%% file: chains.tex
\documentclass[cernpreprint,texlive=2014,txfonts,USenglish,texmf]{atlasdoc}
\usepackage[subfigure]{atlaspackage}
\usepackage{atlasphysics}

\graphicspath{{logos/}{figures/}}

\AtlasTitle{ Study of ordered hadron chains with the ATLAS detector }

\AtlasAbstract{%
The analysis of the momentum difference between charged hadrons in high-energy proton--proton collisions
is performed in order to study coherent particle production. The observed correlation pattern
agrees with a model of a helical QCD string fragmenting into a chain of ground-state hadrons.
A threshold momentum difference in the production of adjacent pairs of charged hadrons is observed,
in agreement with model predictions. The presence of low-mass hadron
chains also explains the
emergence of charge-combination-dependent two-particle correlations commonly attributed to
Bose--Einstein interference. 
The data sample consists of 190 $\mu{\mathrm b}^{-1}$ of minimum bias events collected with 
  proton--proton collisions at a center-of-mass energy $\sqrt{s} = 7$ TeV in the early
 low-luminosity data taking with the ATLAS detector at the LHC.    
}

\author{The ATLAS Collaboration}

\AtlasRefCode{STDM-2014-08}

\PreprintIdNumber{CERN-EP-2017-092}

\AtlasJournal{Phys.\ Rev.\ D}

\hypersetup{pdftitle={ATLAS document},pdfauthor={The ATLAS Collaboration}}

\begin{document}

\maketitle

\tableofcontents

\section{Introduction}

  Studies of correlated hadron production are an important source of information about the early stages of hadron formation,  not yet understood from the theory of strong interactions.  Although experimental high-energy physics employs several phenomenological
 models of hadronization that describe the formation of jets with remarkable accuracy, correlation phenomena are
 more elusive. In particular, the observed excess of nearby equally charged hadrons 
 -- commonly attributed to Bose--Einstein interference -- has never been
 satisfactorily reproduced by Monte Carlo (MC)  models, despite several decades
 of intensive measurements. Furthermore, dedicated studies of these correlations in $WW$ production at
 LEP2 did not confirm the expected presence of correlations between hadrons
 originating from different color-singlet sources~\cite{LEP-2}. 

   Recently, it was pointed out that correlations between like-sign hadrons arise in 
  the causality-respecting model of quantized fragmentation of a
  three-dimensional QCD string~\cite{helix2}, as a consequence of coherent hadron emission~\cite{lund_helixm}.
  The topology of the string and the causal constraint implemented in this model define the 
  mass spectrum and the correlation pattern of emitted hadrons.
  This analysis investigates observables sensitive to predictions of
  the quantized string model. The experimental technique is focused on the
  extraction of a signal from correlated hadron pairs and triplets.    
   
  The paper is organized as follows: Section~\ref{sec:pheno} contains a brief
  overview of phenomenological aspects of the quantized three-dimensional QCD
  string. Section~\ref{observables} recounts the observable features of the model and outlines the strategy
  of the analysis. Section~\ref{sec:detector} describes the ATLAS detector. 
  The data selection and MC event generators are described in Section~\ref{sec:data}. 
 Section~\ref{sec:raw} shows the measured data. Correction of the data to the
 particle level is described in Section~\ref{sec:correction}.
 Section~\ref{sec:results} contains the results and the studies of systematic uncertainties.
 Section~\ref{interpretation} is devoted to interpretation of results and Section~\ref{sec:conclusions} contains concluding remarks.

 \section{Phenomenology of QCD string fragmentation} \label{sec:pheno}

 The Lund string fragmentation model~\cite{lund}, which is
 implemented in the PYTHIA event generator~\cite{PYTHIA8},   uses a
 one-dimensional string to model the QCD confinement.  The string
 is broken randomly by the production of a new quark--antiquark pair (or a pair of
 diquarks if baryons are to be produced). 
 Hadron four-momenta are determined by the relative position and timing of adjacent breakup
 vertices. Hadrons sharing a common breakup vertex are called adjacent hadrons.
 The model imposes a space-like distance between the
 vertices in order to produce hadrons with a positive (physical)
 mass. Despite the absence of a causal connection
 between vertices, the adjacent string breakups cannot be treated as
 random because they define the mass of the created hadron. The mass spectrum
 is enforced in the model by adding the mass constraint to the
 kinematics of the string decay, using hadron masses and widths
 as external parameters. 
 The model relies on the concept of quantum tunneling
 to generate the intrinsic transverse momenta of hadrons: the partons
 created in the string breakup are assigned a transverse momentum with
 a constant azimuthal distribution and with a magnitude drawn randomly according to
 a Gaussian distribution with a tunable width. Local charge and momentum conservation hold
 in the breakup vertex but according to the model, there are no correlations between
 non-adjacent hadrons in the string's transverse plane. 

   The one-dimensional string
 serves as an approximation for a more complex QCD field shape, which may be similar to
 a thin vortex of a type-II superconductor.  The possibility of understanding
 the shape of a QCD string in three-dimensions was first
 studied in Ref.~\cite{lund_helixm}
 with the goal of investigating effects stabilizing the end of the parton shower
 cascade. On the basis of angular properties of gluon
 emission under helicity conservation, the authors of Ref.~\cite{lund_helixm} concluded that
 collinear gluon emissions are absent. On the basis of optimal packing
 of soft non-collinear gluon emissions, it was deduced that the shape of the QCD string
 should be helix-like.
 
  The fragmentation in the transverse plane changes substantially when
  a one-dimensional string is
  replaced by a three-dimensional string and quantum tunneling is
  replaced by gluon splitting into a quark--antiquark pair with
  negligible momentum in the rest frame of the string stretched
  between the color-connected partons.
  Fragmentation of such a string generates intrinsic transverse momentum that
  depends on the folding of the string and implies azimuthal 
  correlations between hadrons.  Azimuthal correlations
  compatible with the helical shape of the QCD string have been
  observed by ATLAS~\cite{STDM-2011-20}. 

\begin{figure}[hbt!]
\includegraphics[width=0.4\textwidth]{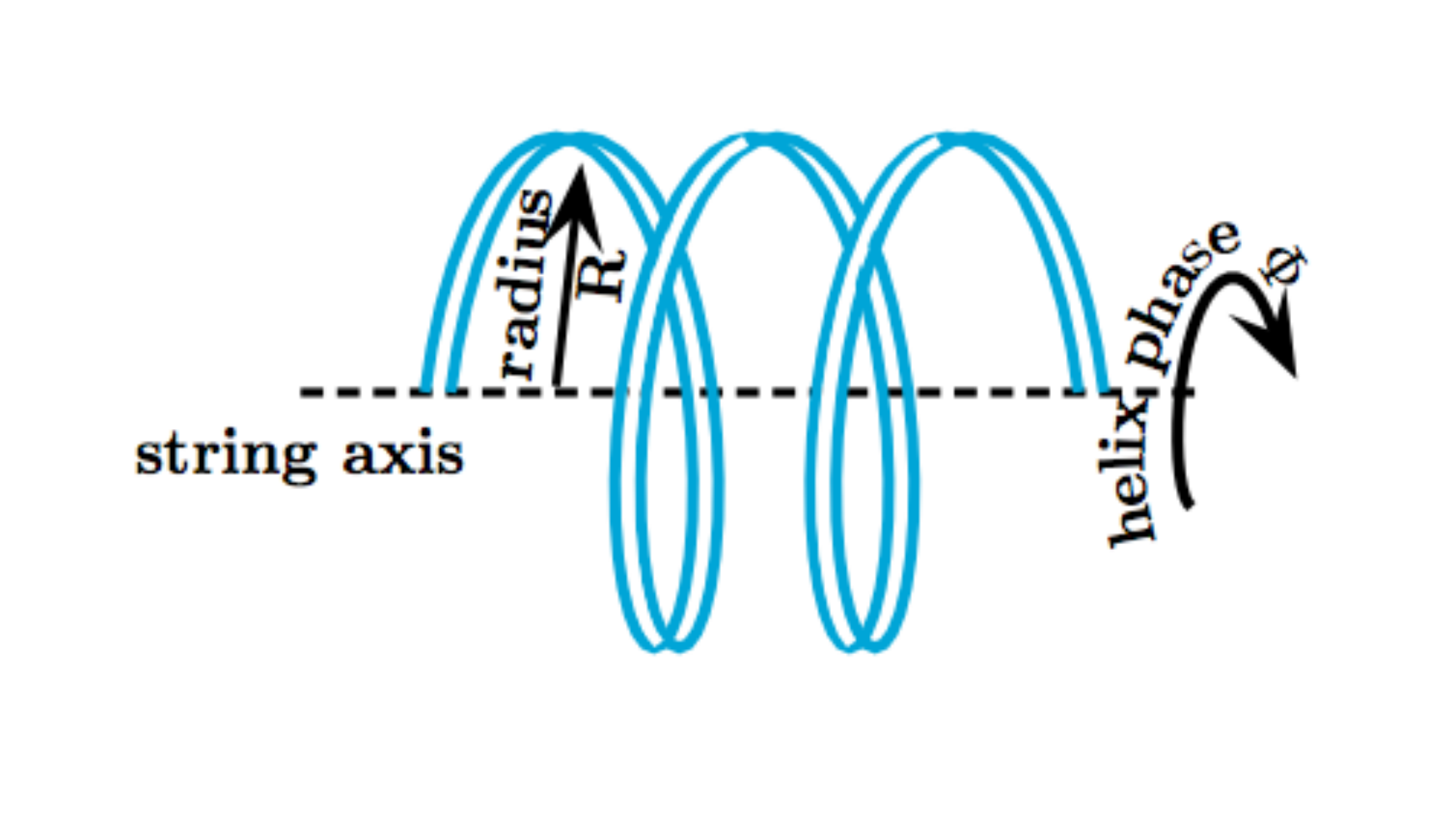}
\includegraphics[width=0.58\textwidth]{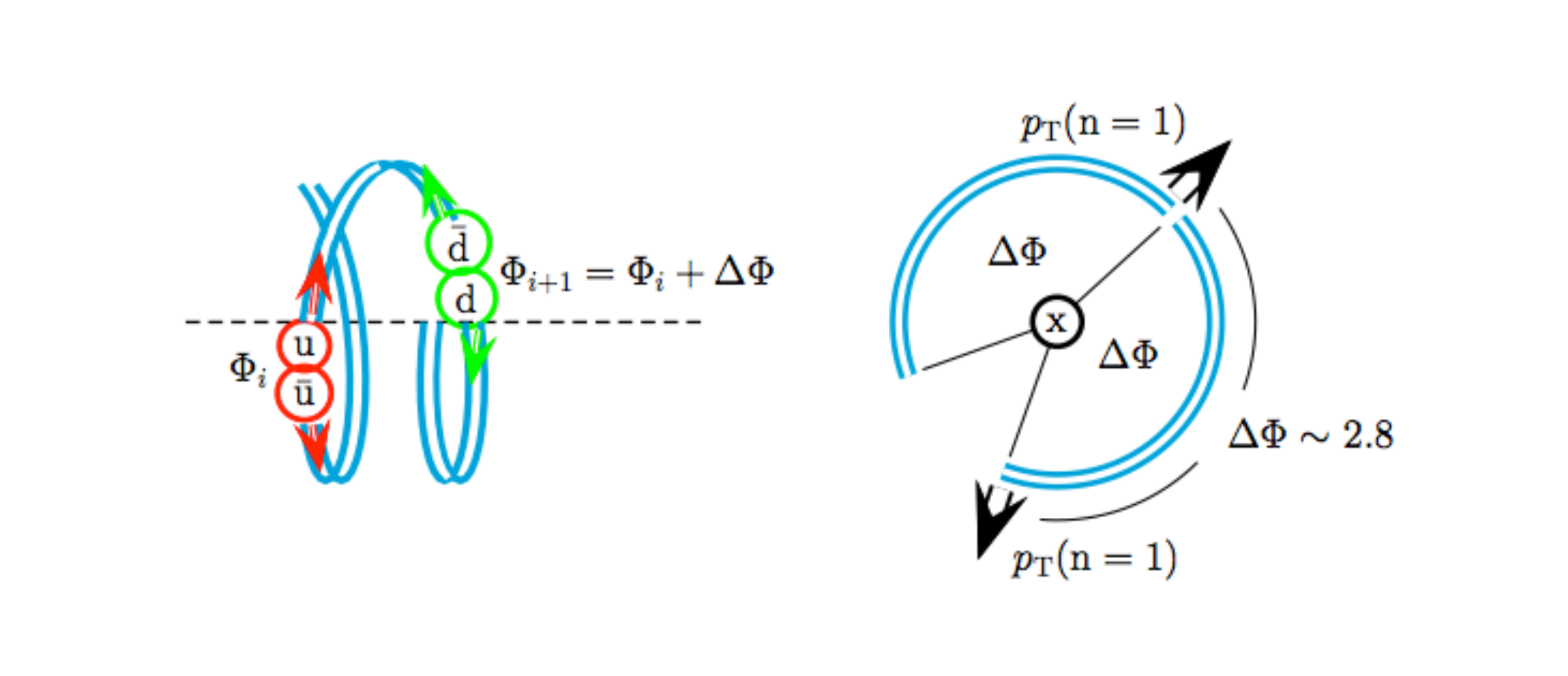}
\caption{  Left: Parameterization of the helical shape of the QCD string.  
Middle: In quantized string fragmentation, the breakup points are separated by the quantized helix phase difference n $\Delta\Phi$, n=1 for the ground-state pion. 
Right:  The shape of the QCD string is reflected in the momentum
distribution of emitted hadrons. The intrinsic transverse momentum of hadrons $p_{\mathrm T}(n)$ is quantized (see Appendix \ref{sec:appendix}). 
The azimuthal angle between intrinsic transverse momenta
of adjacent ground-state pions is equal to $\Delta\Phi$. 
\label{fig:string}}
\end{figure}  
  A fragmentation model working with a three-dimensional string 
 enables cross-talk between break-up vertices to be introduced. When the causal constraint is
  imposed on the fragmentation of a helical QCD string described by radius $R$ and phase $\Phi$ (Figure~\ref{fig:string}), the mass spectrum of light mesons is reproduced by a string breaking in regular $\Delta\Phi$ intervals. A fit of the mass spectrum of pseudoscalar mesons indicates a rather narrow radius of the helical string
 ($\kappa R = 68 \pm 2$~MeV, where $\kappa \sim$~1~GeV/fm is the string tension)
 and a quantized phase difference $\Delta\Phi = 2.82 \pm 0.06$~\cite{helix2}.     

   The effective quantization of the string fragmentation predicts correlations between pairs of hadrons produced along the string, as a function of their rank difference
 $r$.~\footnote{The rank refers to the ordering of hadrons along the
   string; adjacent pairs have rank difference 1.} Correlations can be studied with help of the momentum difference $Q$:
\begin{equation}
   Q_{ij} = \sqrt{ - (p_{i} - p_{j})^2 },
\end{equation}
  where $p_{i},p_{j}$ stand for the four-momenta of particles forming the pair. The numerical values of the predicted momentum difference
 separating pairs of ground-state pions~\footnote{The term ground-state pion denotes the lightest hadron state formed by a string piece with a helix phase difference $\Delta\Phi$, with a causal relation imposed on the endpoint breakup vertices.} with rank differences up to 5 are given in Table~\ref{tab:qrank}. Predictions are calculated in the limit of a locally homogeneous string field with regular helix winding
 which implies vanishing longitudinal momentum difference between pions in the chain.
  Adjacent
 pions are produced with an intrinsic transverse momentum difference of $\sim$266~MeV, which can be seen
 as a quantum threshold for the production of adjacent hadrons. In a chain of adjacent charged pions,  local charge conservation allows for the production of pairs of pions with equal charge for even rank differences
($r$ = 2, 4, ...) and opposite charge for odd rank differences ($r$ = 1, 3, ...) only.
The low-$Q$ region ($Q<$100 MeV) is populated by pairs with $r$ = 2.

    Within the model, a chain of $n$ adjacent ground-state pions has the smallest possible mass for a chain of $n$ adjacent hadrons.
 It can be calculated using the relation 
\begin{equation}
m_{n} = \sqrt{  {n}^2 m_{\pi}^2 + \sum_{i\neq j} Q_{ij}^2 }  ,
\label{eqn:mchain}
\end{equation}
  where $m_{\pi} = m_{n = 1}$ is the pion mass, and $Q_{ij}$  stands for the momentum difference
 of pairs of hadrons forming the chain. Further information about the calculation of model predictions is provided in Appendix~\ref{sec:appendix}.

\begin{table}[t!]{
\caption{ The expected momentum difference between hadrons formed by fragmentation of a homogeneous
  string into a chain of ground-state pions, in the quantized helix string model (see Appendix~\ref{sec:appendix}).
 The 3\% uncertainty is derived from the precision of the fit of the mass spectrum of light
pseudoscalar mesons ($\pi,\eta,\eta'$) \protect\cite{helix2}.\label{tab:qrank}}
\begin{center}
\begin{tabular}{|l|c|c|c|c|c|}
\hline
  Pair rank difference $r$&   1  &  2 & 3 &  4 &  5   \\
\hline
  $Q$ expected  $[{\mathrm MeV}]$  &    266 $\pm$ 8 &  91 $\pm$ 3    & 236 $\pm$ 7  & 171 $\pm$ 5 & 178 $\pm$ 5  \\ 
\hline
\end{tabular}
\end{center}}
\end{table}      

\section{Observable quantum properties of string fragmentation } \label{observables}      

   The analysis uses the two-particle correlations measured for like-sign and opposite-sign hadron pairs
   to study the momentum difference between adjacent hadrons. The possible 
   connection between the enhanced production of
   equally charged pions at low $Q$ and the production of chains of adjacent ground-state
   pions is investigated.  For that purpose, correlations are measured for a 
   selection of exclusive hadron triplets designed to
   isolate the source of correlations (see Section~\ref{chain_selection}) 
  and compared in detail to the inclusively measured two-particle correlations
  \cite{STDM-2013-01,ATLAS-BE-data}. 
   The correlation function is
   defined in a way that facilitates the measurement of adjacent
   hadron pairs (see Section~\ref{correlation_function}).

\subsection{Analysis strategy: chain selection} 
\label{chain_selection} 

    The shortest chain of hadrons from which the properties of a helix
    can be inferred experimentally  is a chain of three charged
    hadrons ($+-+,-+-$), labeled 3h. 
   For the chain selection, it is sufficient to consider only qualitative predictions of the model.
  It is experimentally impossible to reconstruct the history of string
 fragmentation from the momenta of final-state particles only, since the rank difference
 of any given pair of hadrons is unknown a priori.  However, according to the model,
 a chain of ground-state pions will have the lowest possible mass as compared to a chain of arbitrary hadrons,
 and the smallest momentum difference within the chain of charged ground-state pions should be carried by the pair
 of like-sign pions. Therefore, the  chain selection procedure is
 defined, event by event, in the following way: 
 \begin{enumerate}
 \item {Each measured particle is paired with the  like-sign
     particle}  that minimizes the pair momentum
   difference $Q$ calculated from the measured three-momenta. The pion
   mass is assigned to all particles.~\footnote{Throughout this
   paper, the pion mass is assigned to all charged particles in the
   data and in the MC simulation  to reflect the absence of particle
   identification in the data.}
 \item  {Each pair is supplemented with an oppositely charged particle chosen
     to minimize the triplet mass}.
 The resulting three-particle system, ($+-+$) or ($-+-$), is labeled as chain in the following,
  as the charge-conservation constraint is applied to define the relative ordering of particles.
   The chain selection is further refined in order to avoid
   double-counting of particle pairs. The following criteria -- rooted
   in the underlying physics picture of string fragmentation,
   illustrated in
   Figure~\ref{fig:chainflow} --  are applied in an iterative way, preserving the
   configurations with lowest mass found so far.

\item{ The association of particles is verified, in the order of increasing pair momentum difference.} 
  If a particle is associated
  with more than two different like-sign partners, the two pairs with smallest momentum
  difference are retained and the remaining associations are
  discarded. A new search for a closest like-sign partner is performed
  using the still-available particles. Since the algorithm allows a pair of like-sign particles to be associated
  with two different chains,  each proto-chain is assigned a
  weight of $w_i$ = 0.5 or $w_i$ = 1 accordingly, to prevent the double counting of identical chains.

\item{After completion of chains with opposite-sign hadrons,} 
  the association rate of all opposite-sign pairs in the chain selection is
  verified, in the order of increasing chain mass. According to the string fragmentation picture, a pair of adjacent
  hadrons can be shared by at most two adjacent triplet chains
  (Figure~\ref{fig:chainflow}). In the case where a pair of opposite-sign hadrons belongs to three or
  more selected chains, the two chains with lowest mass are retained and a new
  search for an opposite-sign partner is performed for the other chains. If that search fails,
  the weight of the corresponding chain is set to zero. Zero weight is also assigned to
  incomplete chains if there are not enough particles in the event
  to construct triplets.
\end{enumerate}

  At the end of the procedure, the chain selection
  contains $n_{\mathrm ch}$ chains in an event with $n_{\mathrm ch}$ charged
  particles and some of these chains are effectively eliminated,
  having a zero weight. The requirement for the like-sign pair to carry the smallest momentum difference within the chain is not imposed
  in any way: only $\sim$ 1/3 of selected chains contain such a configuration. 
  Although the chain selection builds on generic properties of
  chains of ground-state hadrons by minimizing both the momentum difference for like-sign pairs and the mass of triplets,
  the numerical predictions of the helical string model are not used in the chain
  selection. For the sake of simplicity, the analysis is restricted to the study of triplet chains only.

\subsection{Analysis strategy: definition of the correlation function}
\label{correlation_function}

   In the picture of the string fragmentation, the number of pairs of adjacent hadrons or pairs of hadrons with
a fixed rank difference  is proportional to the number of charged
particles in the sample, while the total number of particle pairs grows quadratically with particle multiplicity.
The choice of the correlation function is therefore driven by the need
to separate the signal from adjacent hadron pairs and the large combinatorial background. 
   
 In the fragmentation of a QCD string, the creation of adjacent like-sign pairs is forbidden by local charge conservation.
For higher rank differences, the like-sign and unlike-sign pairs should be produced in equal amounts due to the random production 
of neutral hadrons in the chain.
It follows that the subtraction of like-sign pairs from the opposite-sign
pairs is a suitable technique for isolating the signal
from adjacent hadron pairs.
The inclusive correlations are
therefore assessed by the correlation function 
\begin{equation}
\Delta(Q)  =  \frac{1}{N_{\mathrm ch}}  \ [  \  N(Q)^{\mathrm OS}-N(Q)^{\mathrm LS} \ ]~,\\
\label{eqn:delta}
\end{equation}
where $N_{\mathrm ch}$ stands for the number of charged particles in the sample, and $N(Q)^{\mathrm OS}(N(Q)^{\mathrm LS})$ denotes the inclusive spectrum of opposite-sign (like-sign) pairs in the sample:
\begin{eqnarray} \label{eqn:inclusive} 
 N(Q)^{\mathrm OS} & = & \sum_{k=1}^{N_{\mathrm ev}}\sum_{i,j=1, \ i\neq j}^{n_{\mathrm ch}^{k}} \delta( q_{i} + q_{j} ) \ \delta( Q - Q_{ij} ), \nonumber \\   
 N(Q)^{\mathrm LS} & = & \sum_{k=1}^{N_{\mathrm ev}}\sum_{i,j=1, \ i\neq j}^{n_{\mathrm ch}^{k}} \delta( q_{i} - q_{j} ) \ \delta( Q - Q_{ij} ).     
\end{eqnarray}
The $\delta(x)= 1-{\mathrm sign}^2(x)$ corresponds to the Kronecker delta with a continuous argument,
which is 1 for $x$=0 and zero otherwise.  $N_{\mathrm ev}$ stands for the number of events, $n_{\mathrm ch}^k$ is the number of
charged particles in event $k$ and $q_{i}$ is the charge of particle $i$.  
 The integral of the $\Delta(Q)$-distribution depends
 on the distribution of the total event charge only, with an upper limit
 of 0.5 for a sample of events with equal number of negatively- and positively-charged particles. Experimentally, the restricted reconstruction acceptance
 creates a charge imbalance which implies a larger loss of opposite-sign pairs compared to that of like-sign pairs 
   and hence diminishes the integral of $\Delta(Q)$ distribution.

   The properties of the correlation function were verified on MC
   samples where no correlations are introduced beyond local charge  and momentum conservation in the string breakup.
   The subtraction of inclusive pair distributions provides the same
   result as the extraction of true adjacent pairs, up to the uncertainty in the particle ordering generated by the
  presence of resonances decaying into three and more charged hadrons (see Figure~\ref{delta_mc_incl} for illustration).
 This implies that the definition of the correlation function to a
 large extent compensates for not knowing  the exact hadron ordering in the
 string fragmentation. Traditionally, correlation studies employ a ratio of $Q$ distributions
  rather than a difference, assuming incoherent or collective
  effects. Such an approach, however, does not eliminate the
  combinatorial background from the measurement and therefore it is far less
  suitable for the measurement of the hadronization effects. 
  A comparison of the two approaches is discussed in Section~\ref{be}.

\begin{figure}[bt!]
\begin{center}
\subfigure[ ]{
  \label{triplet_a}
\includegraphics[width=0.32\textwidth]{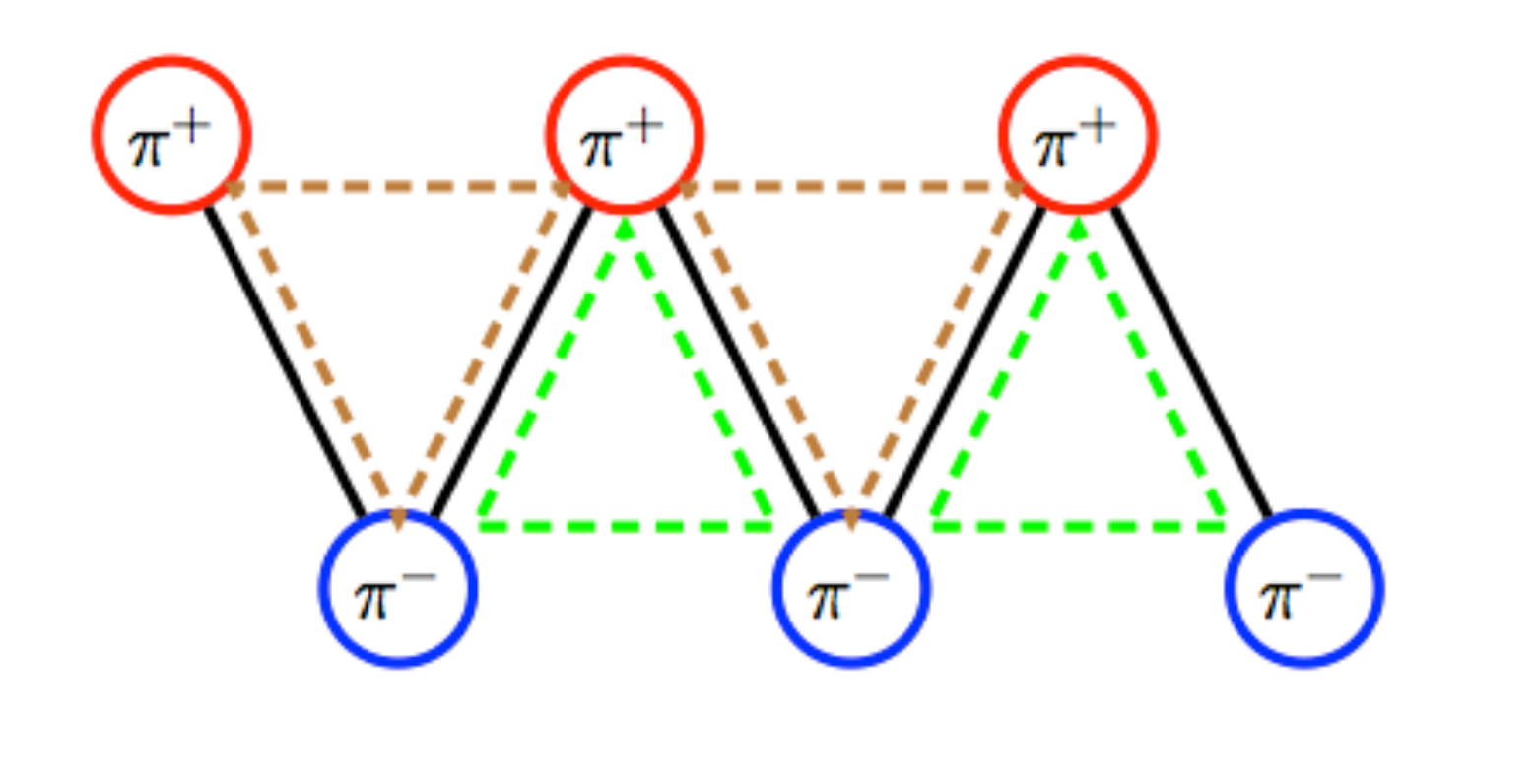}
}
\subfigure[ ]{
  \label{triplet_b}
\includegraphics[width=0.46\textwidth]{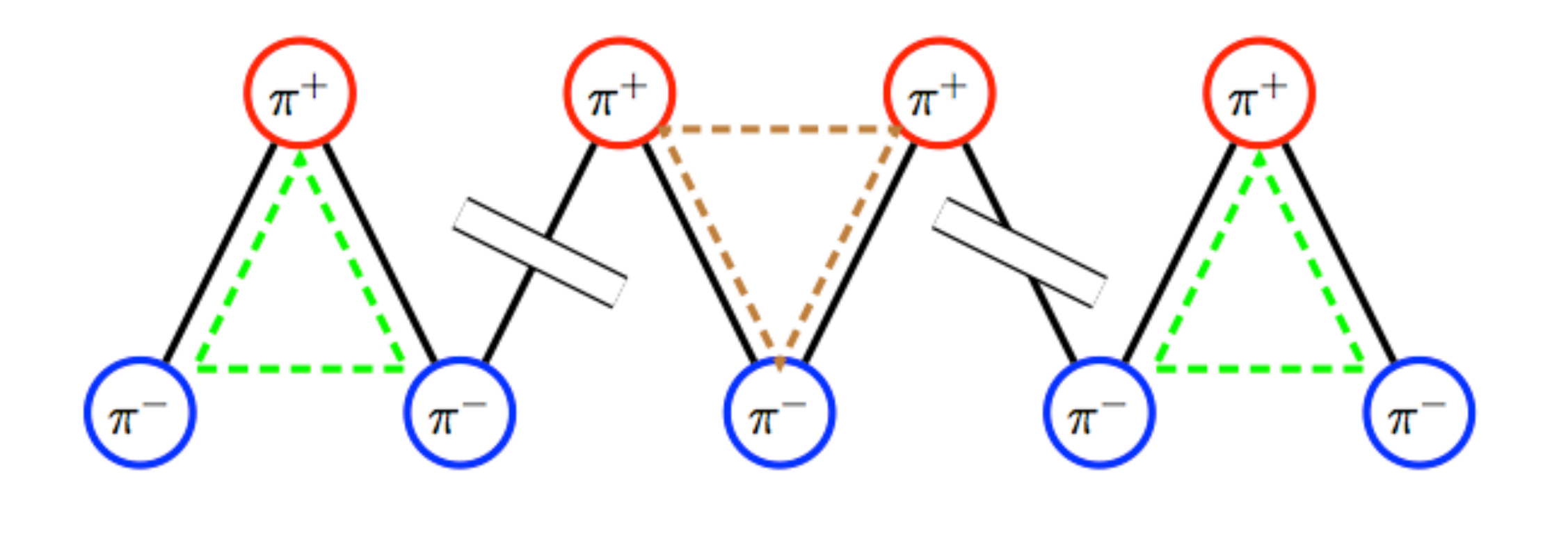}
}
\caption{ A schematic view of a string fragmentation into (a) a long uninterrupted
  chain and (b) disconnected three-hadron chains. The circles
  represent charged hadrons and the black lines indicate the ordering according to the string
  fragmentation history (they connect adjacent hadron pairs). Dashed triangles indicate the triplet chains.
  In the long uninterrupted chain (a),  neighboring triplets share a common
  opposite-sign pair.   
\label{fig:chainflow}}
\end{center}
\end{figure}

  In direct correspondence to Eq.~(\ref{eqn:delta}) the correlations carried by exclusive three-hadron chains
  can be expressed as a sum over contributions from all chains with non-zero weight
\begin{equation}
 \Delta_{\mathrm 3h}(Q) = \frac{1}{N_{\mathrm ch}}  \sum_{k=1}^{N_{\mathrm ev}}\sum_{i=1}^{n_{\mathrm ch}^{k}} w_i \left\{ \ \frac{1}{2} \ \delta(Q-Q_{01}^{i}) +  \frac{1}{2} \ \delta(Q-Q_{12}^{i}) \
    -  \delta(Q-Q_{02}^{i}) \  \right\},
\label{eqn:delta3h}
\end{equation}
where each chain contributes with three entries: two for opposite-sign pairs
at $Q_{01},Q_{12}$ and one for the like-sign pair at $Q_{02}$ (the indexes reflect charge-ordering of particles in the chain). 
 The $w_i$ stand for the weight factor of the $i$-th chain in the event. 

   The scaling of the opposite-sign pair contribution by 1/2 in Eq.~(\ref{eqn:delta3h}) is required
   for proper subtraction of random combinations; physics-wise,
   it corresponds to a hypothesis of an
   uninterrupted chain of charged hadrons 
  where neighboring triplets share an opposite-sign pair, Figure~\ref{triplet_a}. 
  The estimate for disconnected triplets (Figure~\ref{triplet_b}),
  where the opposite-sign pairs are not shared and should be counted with weight 1
  can be obtained from the measurement of $\Delta_{\mathrm 3h}$ (after subtraction of random combinations), by rescaling the opposite-sign pair
  contribution  -- the positive part of the $\Delta_{\mathrm 3h}(Q)$ spectrum -- by a factor of two.
 
\subsection{Analysis strategy: three-body decay}
 
 Quantized fragmentation of the helical string into a chain of charged
 pions is expected to produce a distinct three-body decay pattern. The chain members are separated by a momentum difference that depends 
on their rank difference (Table~\ref{tab:qrank}).
 For a triplet chain, such a signal can be studied with the help of a Dalitz plot.  
  In analogy with studies of $\eta$ decay~\cite{KLOE,WASA}, the Dalitz plot coordinates ($X$,$Y$) are defined as   
 \begin{equation}
     X = \sqrt{3} \  \frac{T_{0} - T_{2}}{\Sigma_{i=0}^2 T_{i}},~~~~~ \ \   Y  =  \frac{3 T_{1}}{\Sigma_{i=0}^{2} T_{i}} - 1,  \\
 \label{eqn:dalitz}
 \end{equation}
   where $T_{i}$ denotes the kinetic energy $E_i(m_\pi)-m_\pi$  of charge ordered particles in the rest frame of the chain (particles 0 and 2
  form the like-sign pair). The method of calculation of predictions of the helix-string model can be found in Appendix~\ref{sec:appendix} (Eq.(\ref{eqn:q})).

\section{ATLAS detector} \label{sec:detector}

The ATLAS detector \cite{ATLAS}
covers almost the entire solid angle around the 
collision point with layers of tracking detectors, calorimeters and muon chambers.   
For the measurements presented in this paper, the trigger system and the tracking devices 
are of particular importance. The following description corresponds to the detector configuration in the first LHC data-taking
period (Run 1).

The ATLAS inner detector has full coverage in $\phi$ and covers the pseudorapidity 
range $|\eta| < 2.5$.\footnote{ATLAS uses a right-handed coordinate system with its origin at the nominal interaction point (IP) in the center of the detector
and  the $z$-axis  along  the  beam  pipe.  The $x$-axis  points  from
the  IP  to  the  center  of  the  LHC  ring,  and  the $y$-axis  points  upward. Cylindrical coordinates $(r,\phi)$ are used in the transverse plane,
$\phi$ being the azimuthal angle around the beam pipe. The
pseudorapidity is defined in terms of the polar angle $\theta$ as
$\eta = -\mathrm ln \ tan(\theta/2) $. } It consists of a silicon pixel detector, a silicon strip detector 
(SCT) and a transition radiation tracker (TRT). These detectors are immersed in a 2~T axial magnetic field. The pixel, SCT and TRT detectors have typical  $r$--$\phi$ position resolutions of 10, 17 and 130 $\mu$m, respectively,  and  the pixel and SCT detectors have $r$--$z$ position resolutions of 115 and 580 $\mu$m, respectively. 
A track traversing the full radial extent would typically have 3 silicon pixel hits, 8 or more silicon strip hits  and more than 30 TRT hits.

The ATLAS detector has a three-level trigger system:~level~1~(L1), level~2 (L2) and the 
event~filter~(EF).
For this measurement, the L1 trigger relies on the beam pickup timing devices (BPTX) and
the minimum-bias trigger scintillators (MBTS).
The BPTX are composed of electrostatic beam pickups attached to the beam pipe at a 
distance $z = \pm 175$~m from the center of the ATLAS detector.
The MBTS are mounted at each end of the inner detector in front of the endcap calorimeter 
 at $z = \pm 3.56$~m and are segmented into eight sectors in azimuth 
and two rings in pseudorapidity ($2.09 < |\eta| < 2.82$ and $2.82 < |\eta| < 3.84$).
Data were taken for this analysis using the single-arm MBTS trigger, formed from BPTX 
and MBTS L1 trigger signals.
The MBTS trigger was configured to require one hit above threshold from either side of the 
detector. The MBTS trigger efficiency was studied with a separate pre-scaled L1 BPTX trigger, 
filtered to obtain inelastic interactions by inner detector requirements at L2 and the EF \cite{STDM-2010-06}.
        
 \section{Data selection and MC event generators} \label{sec:data}

  Event and track selection are identical to those used in Refs.~\cite{STDM-2013-01} and
  \cite{STDM-2010-06}. The data sample consists of 190 $\mu{\mathrm b}^{-1}$ of minimum-bias events collected with 
  proton--proton collisions at a center-of-mass energy $\sqrt{s} = 7$~TeV in the early 2010 ATLAS data taking with negligible contribution from additional $pp$ collisions in the same bunch crossing.
 Events must
\begin{itemize}
\item{ pass a single arm MBTS trigger},
\item{ have a primary vertex reconstructed with at least two associated tracks each with transverse momentum ($p_{\mathrm T}$)  above 100~MeV},
\item{ not have a second primary vertex reconstructed with more than three tracks},
\item{ have at least two good tracks, as defined below}.
\end{itemize}
A reconstructed track passes the selection if it has
\begin{itemize}
\item{ $p_{\mathrm T}>$ 100~MeV, and lies in the pseudorapidity range $|\eta| <$ 2.5},
\item{ absolute values of transverse and longitudinal impact parameters below 1.5 mm, with respect to the event primary vertex},
\item{ a hit in the first pixel layer when expected, and at least one pixel hit in total},
\item{ at least two (for $p_{\mathrm T}>$100~MeV), four (for $p_{\mathrm T}>$200~MeV), or six (for $p_{\mathrm T}>$300~MeV) SCT hits},
\item{ a fit probability above 0.01 for $p_{\mathrm T}>$10~GeV}.
\end{itemize}  
 
 The sample contains $\sim$10 million events and over 200 million reconstructed tracks.
 The detector effects are evaluated using a PYTHIA6.421~\cite{PYTHIA6} event sample with parameter values from the MC09 
tune~\cite{ATL-PHYS-PUB-2010-002}, fully
simulated~\cite{Geant4} and reconstructed using the standard ATLAS
reconstruction chain~\cite{ATLAS-rec}.  According to MC estimates, the selected set of reconstructed charged particles
 consists of 86\% pions, 9.5\% kaons, 4\% baryons and 0.5\%  leptons, while the fraction of non-primary particles is 2.3\%.        
Primary particles are defined as all particles with lifetime longer than 0.3$ \cdot 10^{-10}$ s originating from the primary interaction
or from subsequent decay of particles with shorter lifetime.
 
 Correlation effects that are studied in the present analysis are
 absent in hadronization models and therefore the analysis does not rely
 on MC predictions. For illustration, the data are compared with a
 representative set of hadronization models including PYTHIA8 (4C tune\cite{4Ctune}), 
 Herwig++~\cite{HERWIG++,UE7tune} and Epos~\cite{EPOS}.

\section{Correlation functions at detector level} \label{sec:raw}

\begin{figure}[bth!]
\begin{center}
\includegraphics[width=0.7\textwidth]{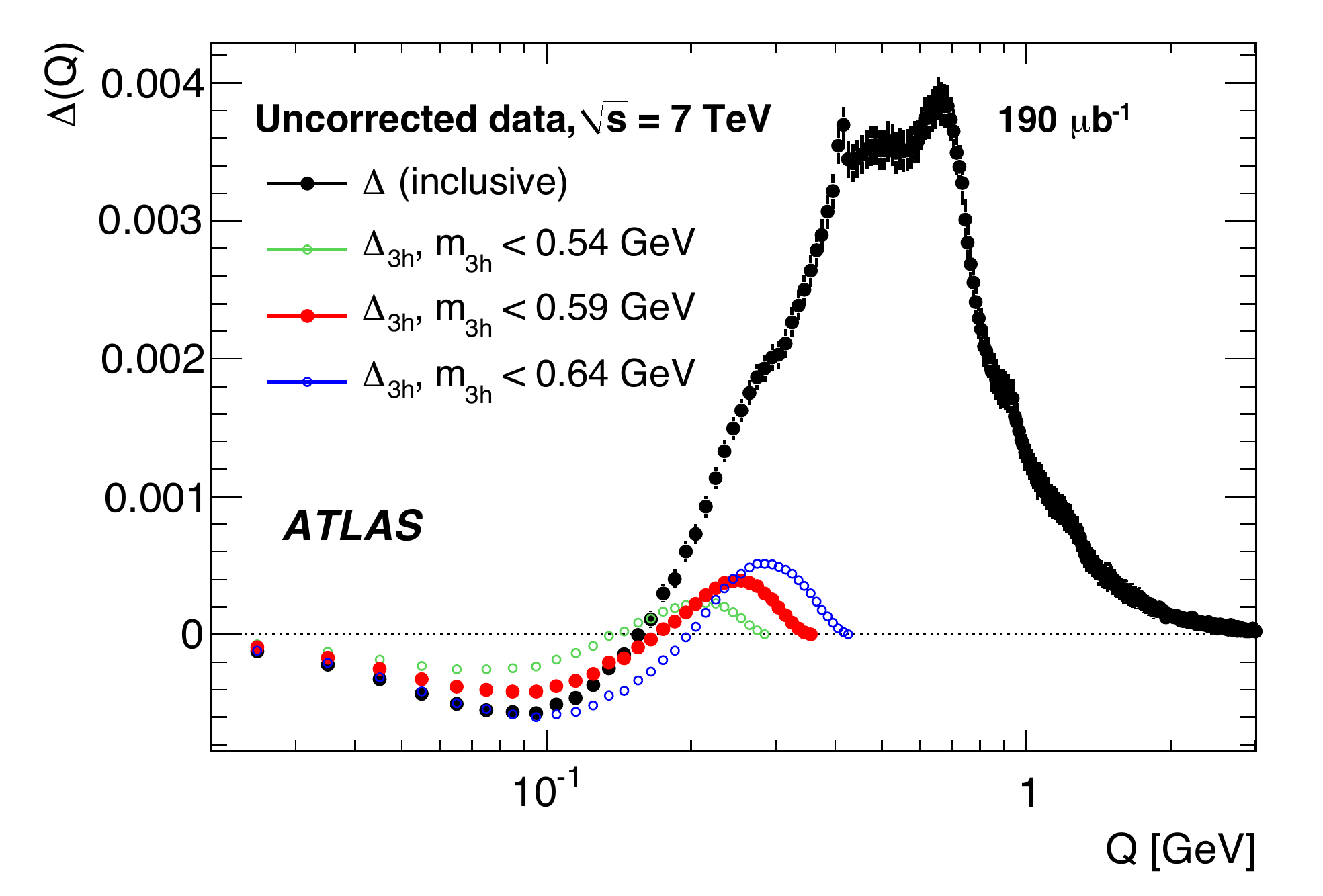}
\caption{ 
 Comparison of  $\Delta_{\mathrm 3h}$ (constructed from pairs belonging to the exclusive three-particle chain selection)
 with $\Delta$ (inclusive two-particle distribution), as a function of the
upper limit on the mass of selected three-hadron chains for uncorrected data.
\label{fig:chain_raw}}
\end{center}
\end{figure}

  The inclusive distribution $\Delta(Q)$ -- as obtained from reconstructed data -- is
  shown in Figure~\ref{fig:chain_raw}.  It shows an enhanced
  production of like-sign hadron pairs at low $Q$, visible 
  as a negative value. 
  The effect is quantified by the  correlation strength (CS) defined as the 
  absolute value of the integral of the negative part of the $\Delta(Q)$ distribution 
\begin{equation}
 CS   = - \int_{\Delta<0} \Delta(Q) \ {\mathrm d}Q. \\
\label{eqn:cs}
\end{equation}

   The correlation function measured for three-hadron chains,
   $\Delta_{\mathrm 3h}(Q)$,  exhibits a shape similar to  $\Delta(Q)$ 
   at low $Q$ (Figure~\ref{fig:chain_raw}). However,  the  shape of the $\Delta_{\mathrm 3h}(Q)$
   also depends on
   the upper mass limit,  $m_{\mathrm 3h}^{\mathrm{cut}}$, imposed on the chain selection. The chain correlation strength (CCS) 
   is defined as
\begin{equation}
 CCS(m_{\mathrm 3h}^{\mathrm{cut}})   = - \int_{\Delta_{\mathrm 3h}<0} \Delta_{\mathrm 3h}(Q) \ {\mathrm d}Q,   \  ~~~~~~~ {\mathrm{for}}  \ {m_{\mathrm 3h}<m_{\mathrm 3h}^{\mathrm{cut}}}. 
\label{eqn:ccs}
\end{equation}      

   A comparison of the two distributions suggests that the selection of
   triplet hadron chains with minimized mass may contain the source of
   two-particle correlations observed in the inclusive sample, but this information is indicative only at this stage due to the absence of 
   unfolding of detector effects that affect the triplet selection more than the inclusive two-particle spectra. The
   average reconstruction efficiency for a pair of charged particles is $\sim 50\%$, for a triplet of charged particles  it drops to $\sim 35\%$.  
   
\section{Correction to particle level} \label{sec:correction}
\label{correction}

\begin{figure}[hbt!]
\subfigure[ ]{
  \label{syst_a}
\includegraphics[width=0.49\textwidth]{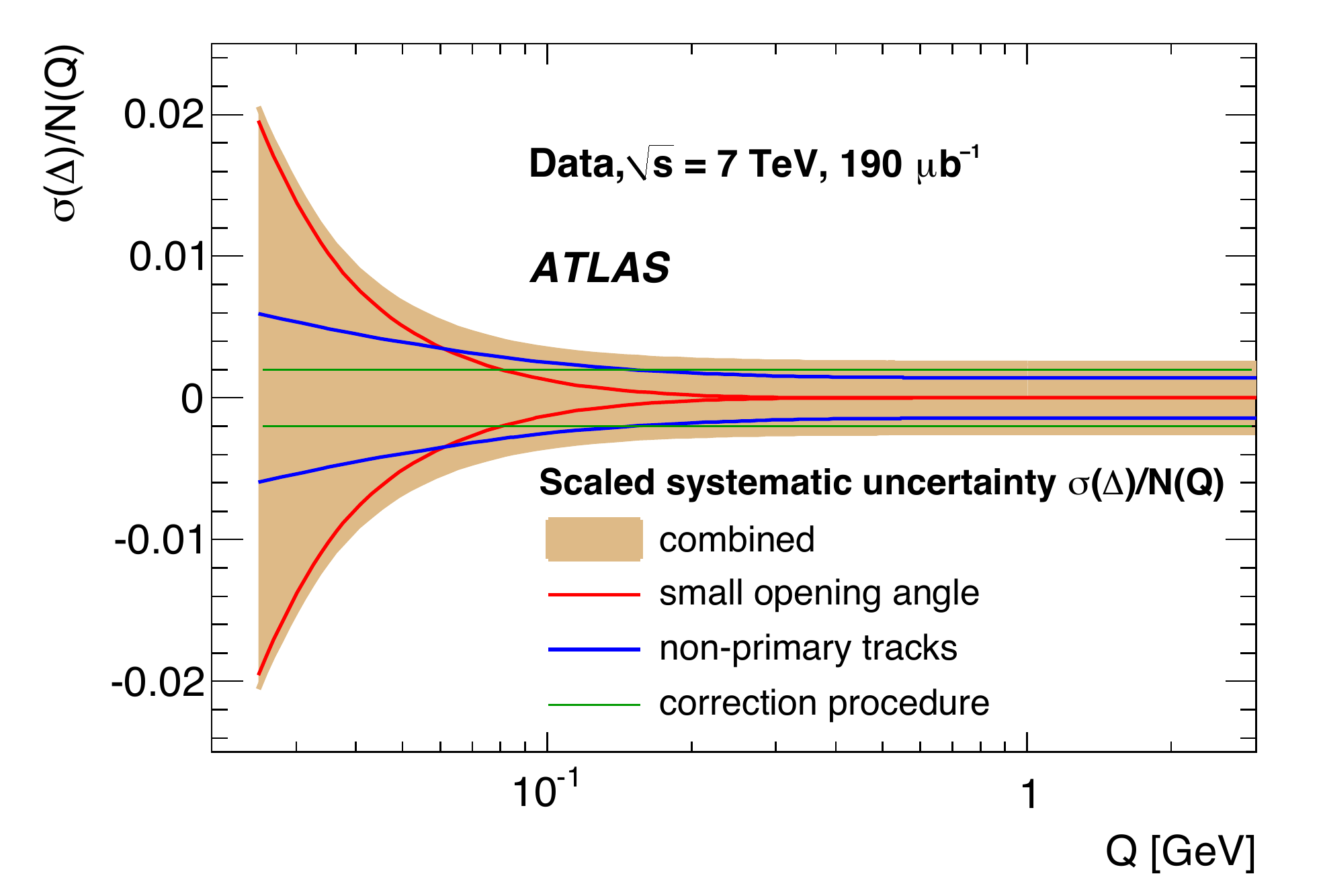}
}
\subfigure[ ]{
  \label{syst_b}
\includegraphics[width=0.49\textwidth]{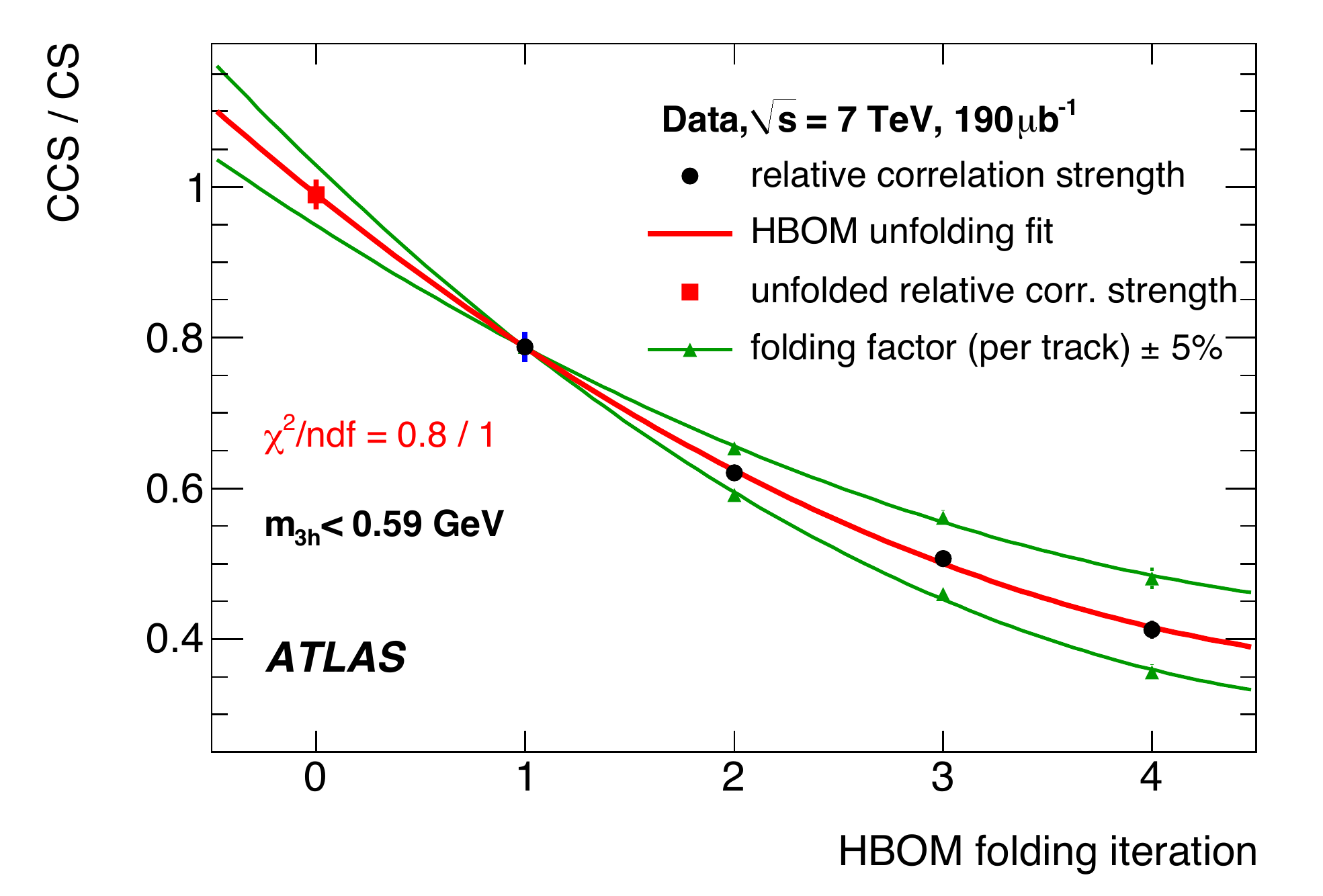}
}
\caption{(a) Components of the systematic uncertainty of $\Delta(Q)$
  related to the reconstruction of pairs of tracks
  normalized to the combined inclusive spectrum $N(Q)= N(Q)^{\mathrm OS}+N(Q)^{\mathrm LS}$. 
 (b) HBOM fit (red curve) which provides the corrected relative correlation strength
 (red square marker). The systematic uncertainty is split into pair
reconstruction uncertainty (error bar superimposed over the raw data point)
and the variation of track folding factors (green lines). 
\label{fig:syst}}
\end{figure}

 The data are corrected for detector effects within the
  acceptance requirements ($p_{\mathrm T}>$100~MeV, $|\eta| <$ 2.5) following the correction procedure established in Ref.~\cite{STDM-2010-06}.
  The track reconstruction inefficiency, the presence of non-primary tracks and the migration of tracks across the acceptance boundaries
  are corrected for by applying
   track-based weights. The vertex and trigger efficiency is corrected for with an event-based weight.
 The dominant component of the uncertainty of the track  weighting factors comes from the dependence
 on the generated MC sample. Anticipating a strong contribution to the measurement from the low-$p_{\mathrm T}$ region, the fully simulated PYTHIA6 sample is 
split into non-diffractive and diffractive components. The former  is used to calculate track-based weights, the latter is used to evaluate the uncertainty of the correction in the low-$p_{\mathrm T}$ region. The observed difference for the inclusive $Q$ spectra (Eq.~\ref{eqn:inclusive}) is $\sim$10\%, without a significant $Q$ dependence. 
  The uncertainty of the track weighting factors due to the imperfect detector description is evaluated using MC samples simulated and reconstructed with a 10\% increase of material in the inner detector, which corresponds to the uncertainty of the detector description (Ref.~\cite{STDM-2010-06}). The observed change of the inclusive $Q$ does not exceed 2\%.
  Both effects are combined and translated into an effective single-track weight uncertainty of 5\%.  

   The study of hadron pairs is sensitive to detector
  effects related to the proximity of reconstructed tracks that
  are not explicitly included in the track-based weights: the reduced reconstruction efficiency for
  pairs of tracks with very low opening angle and the correlated non-primary particle production are taken into
  account via additional correction factors and additional systematic uncertainty.
  Both are parameterized in MC samples as a function of the opening angle between particles and the parameterization is 
  convolved with the opening-angle distribution
  obtained from the data.  The subtraction of the like-sign hadron pair spectrum from the opposite-sign pair spectrum
  implies a large cancellation of these
  effects and cancellation of systematic effects in general.  This renders
  the experimental technique very robust. The uncertainty of track-based weights is effectively removed from the integral of the $\Delta(Q)$ distribution thanks 
  to the appropriate choice of the normalization.  The residual bias of the pair-correction procedure is evaluated in MC samples
  and added to the pair reconstruction systematic uncertainty.  
  Figure~\ref{syst_a} shows the $Q$ dependence of the $\Delta(Q)$ uncertainty related to the combined pair reconstruction 
  uncertainty and its components. The uncertainty is larger than (or comparable to) the correction observed in the MC simulation.
  The corrected inclusive two-particle $Q$ spectra for like-sign and opposite-sign pairs  used in this analysis,
 $ N(Q)^{\mathrm LS}$ and $N(Q)^{\mathrm OS}$,  were published earlier \cite{STDM-2013-01}, and can be obtained from Ref.~\cite{ATLAS-BE-data}.

 The correction of triplet chains needs to take into account the impact of the track reconstruction inefficiency on the
 chain selection algorithm described in Section~\ref{chain_selection}. 
 The correction for this effect 
 is handled by the HBOM correction technique~\cite{hbom}. In the HBOM
 method, the reconstructed tracks are randomly removed from the sample
 according to the detector reconstruction efficiency, which is parameterized in terms of charged-particle $p_{\mathrm T}$ and pseudorapidity.
 Hence the HBOM iteration corresponds to the folding of the data with detector effects. 
 The chain selection is repeated using remaining tracks, for several HBOM folding iterations, and the results are used to
 establish the functional dependency of the measured quantity on detector effects. The parameterization of this dependency
 is then extrapolated to the detector-effect free (zeroth folding iteration) result.  The
 statistical correlations between HBOM folding iterations are suppressed by
 resampling the probability to retain a given track in each folding iteration independently.

 The systematic uncertainty of the HBOM method is extracted directly
 from the data by performing an additional HBOM unfolding to the detector level, where the detector-level data is taken as the reference 
 and compared to the HBOM unfolding based on remaining folding iterations.

   The relative correlation strength CCS/CS is unfolded using the HBOM technique as shown in
 Figure~\ref{syst_b} for a chain mass limit value of 0.59~GeV. The raw data and
 three HBOM folding iterations are fitted using a second-order polynomial.
 Within the CS and CCS integral ranges, statistical uncertainties of CS and CCS are highly correlated and
 therefore the statistical uncertainties of input points are calculated
 assuming maximal overlap between samples. 
 The extrapolation of the fitted function to the zeroth folding iteration yields the
 unfolded result and the statistical error estimate.  
 The systematic uncertainty is split into
 two components indicated on the plot:  the combined pair reconstruction uncertainty
 that is considered fully correlated between HBOM
 iterations and the uncertainty of the folding factors, equal to the uncertainty of track-based weights discussed above. 

\begin{figure}[b!]
\subfigure[ ]{
  \label{hbom_a}
\includegraphics[width=0.48\textwidth]{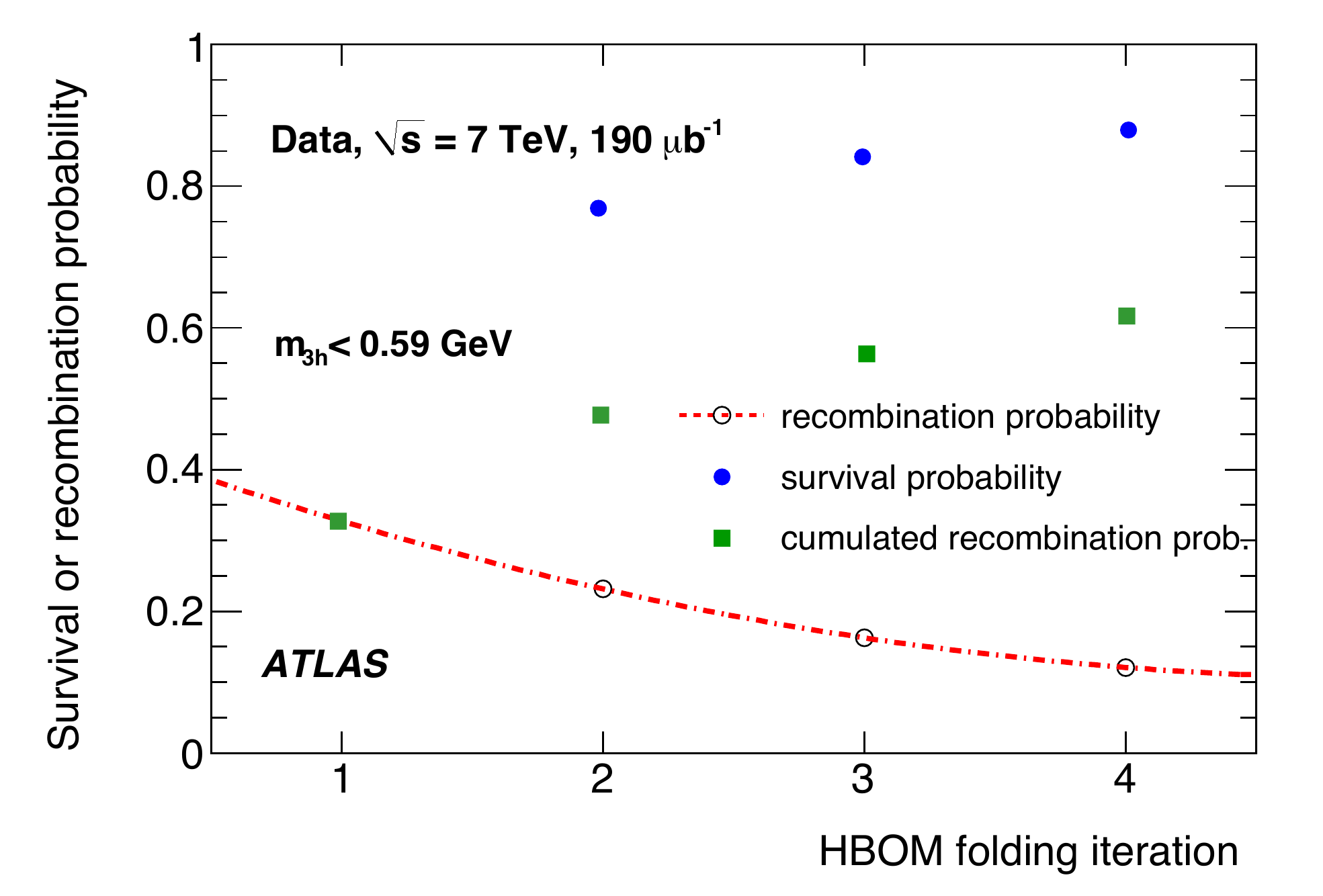}
}
\hspace{0.025\textwidth} 
\subfigure[ ]{
  \label{hbom_b}
\includegraphics[width=0.48\textwidth]{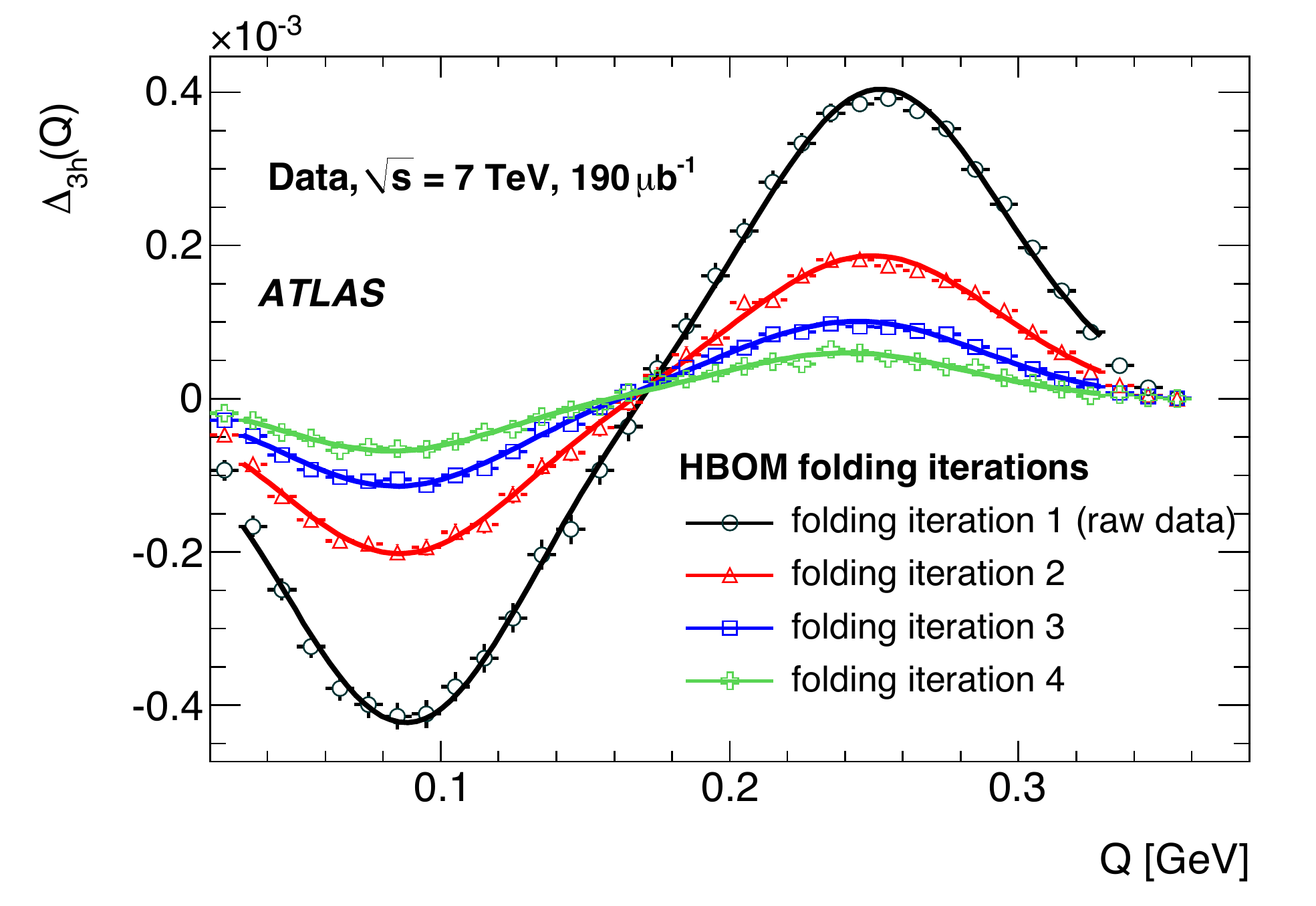}
}
 \caption{ 
Demonstration of HBOM unfolding procedure.
(a) Unfolding fit of the chain recombination probability, used
 to correct the recombination effects in the detector-level chain selection.
The cumulated recombination probability corresponds to the fraction of ``non-original'' chains - those not-existing at the particle level - in each HBOM iteration.
(b)  $\Delta_{\mathrm 3h}$ per
HBOM iteration, for chain mass limit of 0.59~GeV, fitted with a
double-Gaussian parameterization. Both plots show statistical errors only.
 \label{fig:hbom}}
\end{figure}

  The analysis employs two different
 techniques in order to correct the shape of $\Delta_{\mathrm 3h}$. The generic correction
 consists in the evaluation of the fraction and of the shape of chains affected by track recombination.
 A chain which does not loose any track in a given HBOM folding iteration is labeled as a ``surviving'' chain.
 The chain survival probability for the $i$-th HBOM iteration ($i>$1) is calculated as the number of chains surviving
 from the ($i$-1)-th iteration divided by the number of chains selected in the $i$-th iteration.
 Figure~\ref{hbom_a} shows, for a fixed chain mass limit, the chain survival probability as
 measured for three consecutive HBOM folding iterations. The chain recombination probability is complementary to the chain 
 survival probability.
 It serves as an input for the
 unfolding of the recombination probability at the detector level ($\sim
 34\%$ for a chain mass limit of 0.59~GeV). The distribution of $\Delta_{\mathrm 3h}(Q)$ for the
 recombined chains is obtained as a difference between the  $\Delta_{\mathrm 3h}(Q)$ distribution
 obtained in a given folding iteration and the  $\Delta_{\mathrm 3h}(Q)$ distribution of chains
 surviving from the previous iteration. The shape of $\Delta_{\mathrm 3h}(Q)$ for the recombined chains, together with the normalization obtained from  the unfolded recombination rate,  are used to produce an estimate
 of the contribution of recombined chains to the detector-level measurement. After subtraction of the
 recombined chains, the raw data are unfolded using
 track-based weight factors, in analogy with the unfolding of
 inclusive pair spectra.

    The second technique is designed to unfold the parameterized shape of $\Delta_{\mathrm 3h}(Q)$,
 which is well approximated by a sum of two Gaussian functions describing the residual difference between the opposite-sign and like-sign pair content:
\begin{eqnarray}
   \Delta_{3\mathrm{h}}(Q) & = &  f_{\mathrm LS}(Q;Q_{\mathrm LS},\sigma_{\mathrm LS})+ f_{\mathrm OS}(Q;Q_{\mathrm OS},\sigma_{\mathrm OS}) \nonumber   \\ 
        & = &  n_{\mathrm LS} \exp{ \left( \frac{-(Q - Q_{\mathrm LS})^2}{2 \sigma_{\mathrm LS}^2} \right) } +  n_{\mathrm OS} \exp{ \left( \frac{-(Q - Q_{\mathrm OS})^2}{2 \sigma_{\mathrm OS}^2} \right) }, 
\label{eqn:gaus}
\end{eqnarray}
   where $n_{\mathrm LS}<0$ ($n_{\mathrm OS}>0$)  are scale factors, $Q_{\mathrm LS}$($Q_{\mathrm OS}$) indicate the position of Gaussian peaks, and 
   $\sigma_{\mathrm LS}$ ($\sigma_{\mathrm OS}$) correspond to the width of peaks, for like-sign (opposite-sign) pairs.  
   
  The shape is fitted for four folding iterations (including the raw
 data),  and the fit parameters and correlations between them are
 unfolded using the HBOM technique.
  Figure~\ref{hbom_b} shows the fits of the $\Delta_{\mathrm 3h}(Q)$
  distribution for the $m_{\mathrm 3h}<$0.59 GeV, in the fit range
  $Q\in(0.03,0.33)$ GeV. The bin errors of
  fitted distributions are statistical only.

\begin{figure}[htb!]
\begin{center}
\includegraphics[width=0.47\textwidth]{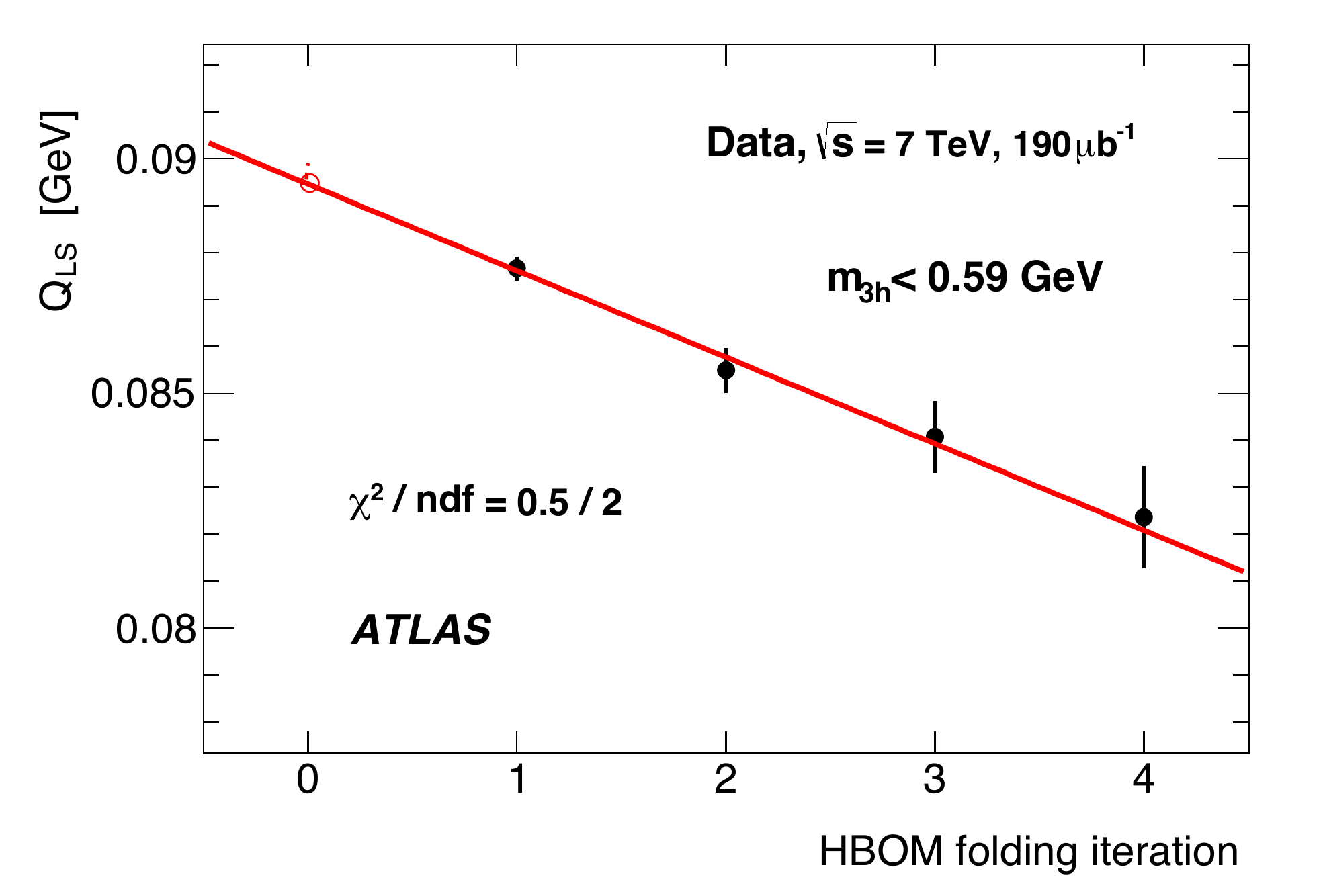}
\includegraphics[width=0.47\textwidth]{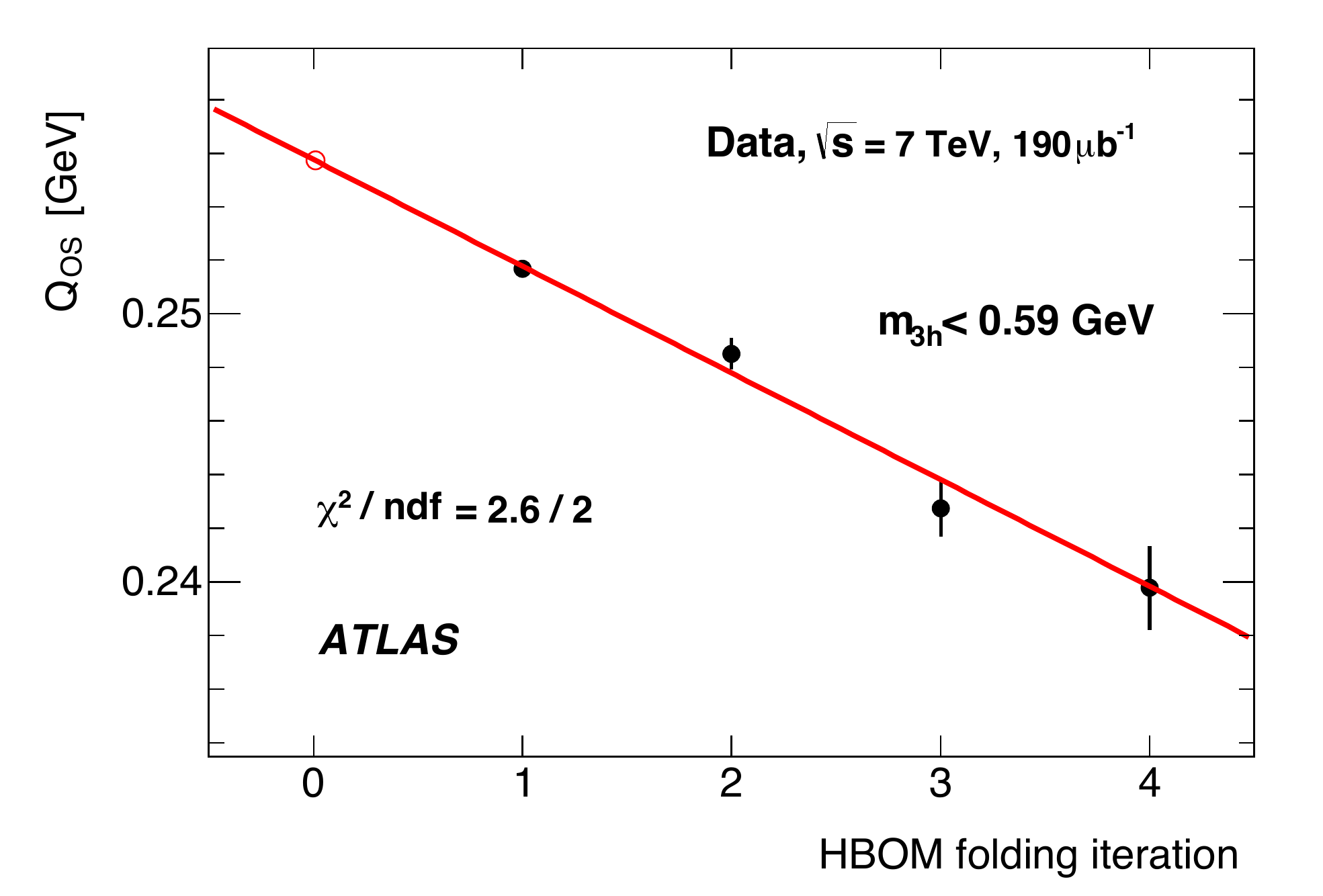}
\includegraphics[width=0.47\textwidth]{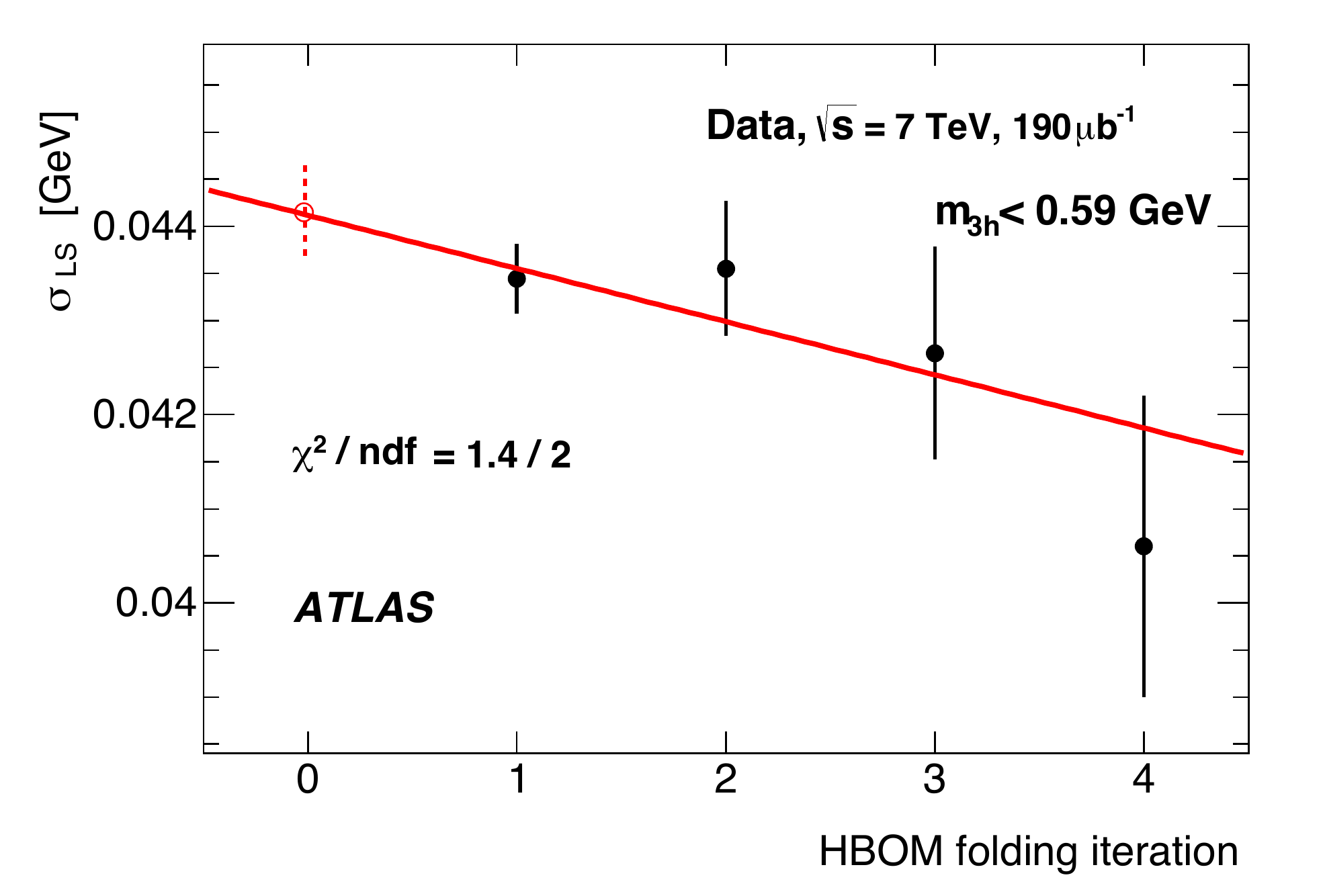}
\includegraphics[width=0.47\textwidth]{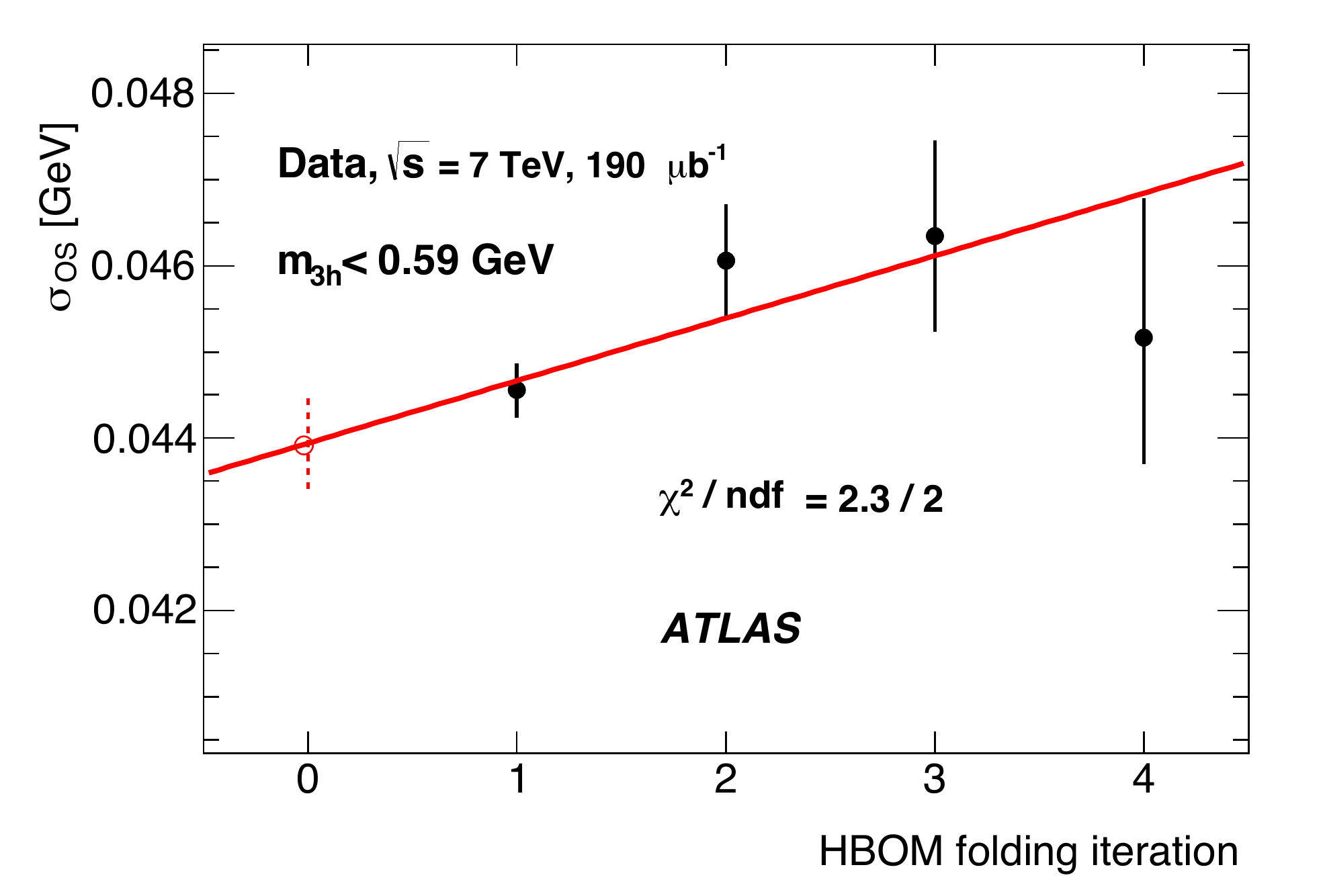}
\caption{ 
 The unfolding of the peak position (top) and of the Gaussian width (bottom),
 for the like-sign pairs (left) and opposite-sign pairs (right),
 for the chain mass limit of 0.59 GeV, using the statistical errors only. The black closed points indicate the values of fit parameter obtained from the fit of $\Delta_{\mathrm 3h}$, the open red point is the unfolded value obtained from the HBOM fit.
\label{fig:qunf_mfix}}
\end{center}
\end{figure}
  Figure~\ref{fig:qunf_mfix} shows the HBOM unfolding extrapolation of the fit parameters: the position and the width of the negative (like-sign pair, LS) peak, and the position and the width of the positive (opposite-sign pair, OS) peak. The stability of the unfolding fit is evaluated using a fit with a polynomial of higher order, and by varying the
 range of the fit of the $\Delta_{\mathrm 3h}(Q)$ distribution.
 Good stability is observed in the unfolding fit of the Gaussian functions
 mean and width values, the results of the fit variations are compatible
 within the statistical errors, and therefore no additional systematic
 uncertainty is attributed to the fit procedure.  It is not necessary
 to unfold the normalization parameters via the HBOM method since they
 are fixed with much higher precision by the adjustment of the
 correlation strength.  The correlation coefficient is  $-0.3$ (0.5) for fitted $Q_{\mathrm LS}$
  and $Q_{\mathrm OS}$ ($\sigma_{\mathrm LS}$ and $\sigma_{\mathrm OS}$) values. 
 
\section{Results} \label{sec:results}

The value of the upper limit on the chain mass for which the chain selection reproduces the
inclusive correlation strength is obtained by interpolation
between HBOM results obtained by variation of the chain mass limit
$m_{\mathrm 3h}^{\mathrm{cut}}$ by 10~MeV (Table~\ref{tab:mfit}).

 The resulting adjusted chain mass-limit value
\begin{equation}   
  m_{\mathrm 3h}^{\mathrm{cut}}(CS=CCS)  =   591 \pm 2 \ {\mathrm(stat) \pm 7 \mathrm (syst)  \ MeV} ,
\label{eqn:mcut}
\end{equation}
  is in agreement with the value $m_{n=3} = 570 \pm 20$~MeV expected
  by the helix fragmentation model, Eq.~(\ref{eqn:mchain}).  

\begin{table}[tb!]{
\caption{ The unfolded relative correlation strength CCS/CS for three values of the upper limit on the mass of triplet chain (column 2).
 Interpolating between the observed values of CCS/CS, the value of  $m_{\mathrm 3h}^{\mathrm{cut}}$ is adjusted to yield CCS/CS=1 (column 3).  
 The systematic uncertainty combines reconstruction uncertainty and the uncertainty of folding factors. \label{tab:mfit}}
\begin{center}
\begin{tabular}{|c|c|c|c|c|}
\hline
 Parameter  &   \multicolumn{3}{c|}{ $m_{\mathrm 3h}^{\mathrm{cut}}$ (input)   }  &     \\
    $[$MeV$]$ &  580   & 590    & 600 &  interpolation \\
 \hline
  CCS/CS $\pm \sigma(\mathrm stat) $  &    0.88 $\pm$ 0.02 &   0.99 $\pm$ 0.02   & 1.09 $\pm$ 0.02  &  1.00  $\pm$
  0.02 (stat) $\pm$ 0.07 (syst)      \\
 \hline
  $m_{\mathrm 3h}^{\mathrm{cut}}$ adjusted  &   \multicolumn{3}{|c|}{}        &   591 $\pm$ 2 (stat) $\pm$ 7 (syst)     \\
\hline 
\end{tabular}
\end{center}}
\end{table}

\begin{figure}[bh!]
  \begin{center}
 \includegraphics[width=0.7\textwidth]{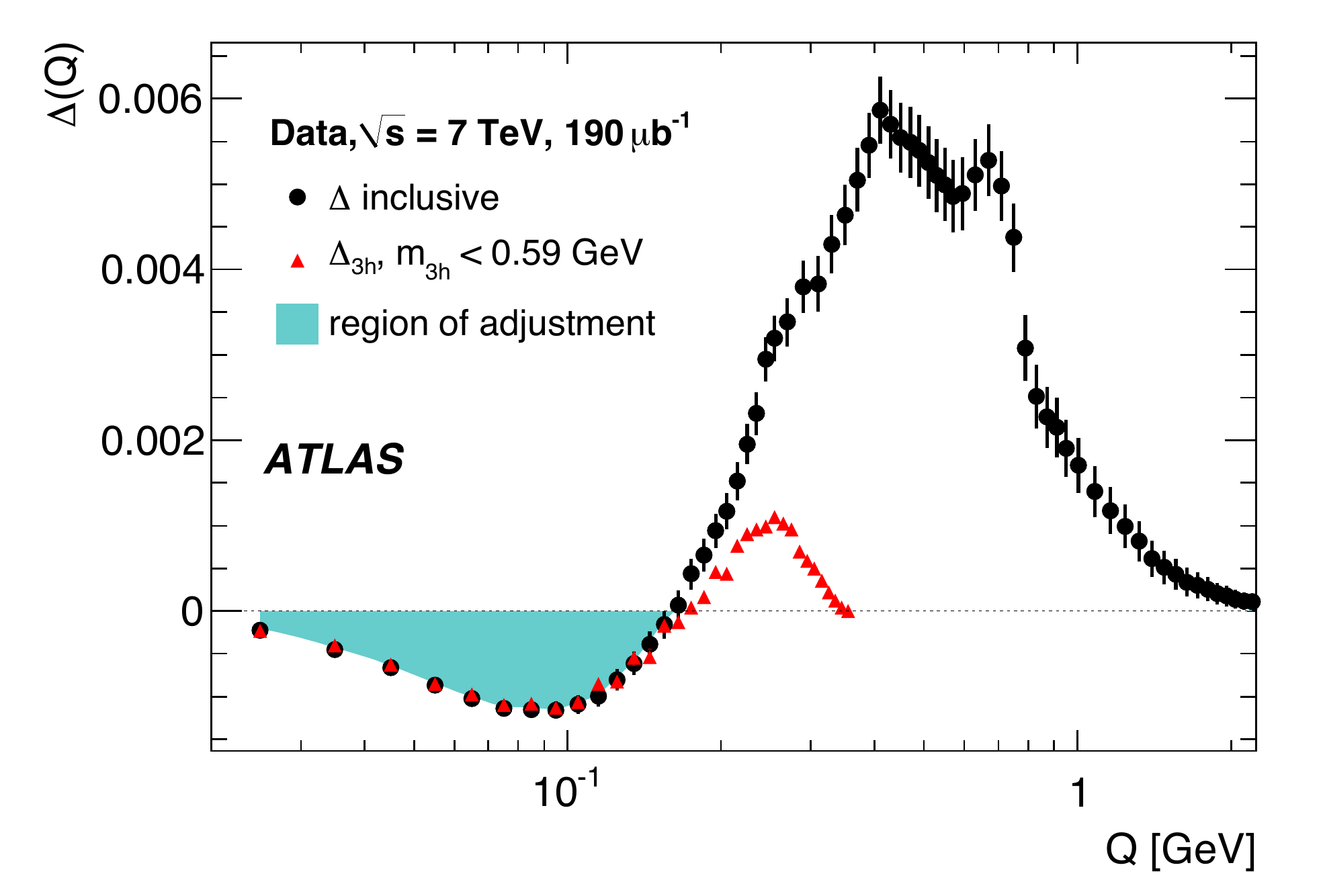}   
 \caption{ 
The corrected $\Delta(Q)$ 
 is compared with the corrected contribution from low-mass three-hadron chains $\Delta_{\mathrm 3h}(Q)$. The chain mass limit is set to a value of
  $m_{\mathrm 3h}^{\mathrm cut}$= 0.59~GeV, which reproduces the excess in the inclusive
 like-sign pair production at low $Q$ (shaded area). Bin errors indicate the combined statistical and reconstruction uncertainty. The uncertainty of the track-based weights is absorbed in the adjustment procedure and translated into uncertainty of the upper chain mass limit.
 \label{fig:mcut}}
\end{center}
\end{figure}  

\begin{table}[ht!]{
\caption{ Unfolded double-Gaussian parameterization of $\Delta_{\mathrm 3h}$
  for a 10~MeV
  scan of the chain mass limit and interpolation to the best estimate of $m_{\mathrm 3h}^{\mathrm{cut}}$
  (see also Eq.~(\ref{eqn:mcut})) compared to the helix model predictions~\cite{helix2}. The systematic uncertainty accounts for all reconstruction effects except
  the uncertainty associated with track-based correction factors, which serves as input for the evaluation of the chains selection uncertainty (Table~\ref{tab:msys}). \label{tab:dqfit}} 
\begin{center}
\begin{tabular}{|c|c|c|c|c|c|}
\hline
 Parameter  &   \multicolumn{4}{c|}{ $m_{\mathrm 3h}^{\mathrm{cut}}$  $[$MeV$]$   }  & QCD helix model    \\
       &   580  & 590    & 600 &  591(best estimate)  &  predictions  $[$MeV$]$ \\
\hline 
  & \multicolumn{3}{c|}{$\pm \sigma(\mathrm stat)$ } & $\pm \sigma(\mathrm  stat) \pm \sigma(\mathrm rec)$  &  \\ 
 \hline
  $Q_{\mathrm LS}$  &       ~~86.6 $\pm$ 0.4    & ~~89.4 $\pm$ 0.4 
  &  ~~92.2 $\pm$ 0.4  &  ~~89.7 $\pm$ 0.4 $\pm$ 1.2  &   ~~91 $\pm$ 3  \\
  $\sigma_{\mathrm LS}$    &      ~~41.4 $\pm$ 0.6     & ~~44.1  $\pm$ 0.6
  &  ~~46.5 $\pm$ 0.7 & ~~44.3 $\pm$ 0.6 $\pm$ 2.0  &  - \\
   $Q_{\mathrm OS}$  &   248.3 $\pm$ 0.5  & 255.8 $\pm$ 0.5 & 262.9 $\pm$ 0.5
 & 256.4 $\pm$ 0.5 $\pm$ 1.8  &  266 $\pm$ 8\\
  $\sigma_{\mathrm OS}$    &     ~~40.9 $\pm$ 0.5    & ~~43.9 $\pm$ 0.6
  & ~~46.5 $\pm$ 0.7 & ~~44.2 $\pm$ 0.6 $\pm$ 1.5  & - \\
\hline 
\end{tabular}
\end{center}}
\end{table}      

  The comparison of the corrected inclusive correlation function
  $\Delta(Q)$ with the corrected correlation function $\Delta_{\mathrm
    3h}(Q)$  representing the contribution from the chain selection
  for the adjusted chain mass limit is shown in Figure~\ref{fig:mcut}.
  The region of adjustment, indicated by the shaded area,  corresponds to the region where the enhanced like-sign pair
  production is observed. 

  In order to measure the shape of the correlation function that corresponds to the adjusted
  chain mass limit, the unfolding of the parameterized shape of $\Delta_{\mathrm 3h}(Q)$ is repeated for several chain mass limit
 values (0.58, 0.59 and 0.6~GeV)  and the unfolded parameters are
 interpolated to the adjusted chains mass limit value (Table~\ref{tab:dqfit}).  The
 reconstruction systematic uncertainty is evaluated by applying a correlated smearing of
 input bin values according to the pair-reconstruction systematic uncertainty.
 The variation of folding factors has negligible impact on the shape
 of the distribution.

 The unfolded $\Delta_{3h}$ measured in the chain selection with adjusted chain mass limit reproduces the shape
  of the inclusive correlation function in the low-$Q$ region.   
 In addition, good agreement is observed between the measurement and the helix model predictions for a chain of three ground-state pions (last column in Table~\ref{tab:dqfit}). 
 The width of the peaks is not predicted by the model, although it can be assimilated with
 the fluctuations of the helix shape of the field. The experimental resolution is better than 10~MeV in the low-$Q$ region.
 
\subsection{Stability of results}

  The analysis was repeated using a single-Gaussian
  parameterization, with independent fits of the negative and
  positive regions of $\Delta_{\mathrm 3h}(Q)$. The change of the fitting function had no significant impact on the results.  
  In addition, the analysis was repeated with HBOM folding accompanied by
  smearing of reconstructed track parameters according to the reconstruction uncertainty. No significant change in the results was observed.

   The stability of the results was verified by a variation of the acceptance requirements:
 the pseudorapidity range was reduced from $|\eta| < $ 2.5 to $|\eta| < $ 1,
 and in a separate study, the transverse momentum threshold was raised from 100 to 200~MeV.
 The study of the impact of  track migration across the acceptance boundary was performed:
 a three-hadron chain containing a particle beyond the acceptance boundary is not reconstructed
 but the remaining pair contributes to the inclusive spectrum. The fraction of lost chains was
 traced and found to be 17\% across the pseudorapidity limit at $|\eta|$ = 1 and 30\%  across the transverse
 momentum limit at $p_{\mathrm T}$ = 200~MeV. Taking into account the track migration, no significant
 differences were observed in the correlation shape $\Delta_{\mathrm 3h}$
 measured within restricted acceptance regions.

  The loss of chains due to a track beyond the detector acceptance, as well as the efficiency of the chain selection cannot
 be fully assessed by the analysis, and the measured chain mass limit is to be considered as an upper limit only.
 An estimate of the mass limit range can be made in the reconstructed data from the rate of
 like-sign pairs at low-$Q$ not associated with
 the chain selection($\sim 30\%$). Under the hypothesis that one third of the  unassociated like-sign pairs
 belong to non-reconstructed chains with very low mass (10\%), the chain mass limit would have to be decreased by $\sim 10$~MeV to keep CS=CCS
 (Table~\ref{tab:mfit}).
 Thus an asymmetric error of $-10$~MeV is added and propagated to the systematic
 uncertainty of the $\Delta_{\mathrm 3h}(Q)$ parameterization. Table~\ref{tab:msys} provides an overview of systematic uncertainties
 related to the chain selection.

\begin{table}[h!]{
\caption{ Overview of systematic uncertainties derived from the
  variation of the chain selection. \label{tab:msys} }
\begin{center}
\begin{tabular}{|c|c|c|c|c|c|c|c|c|}
\hline
 Measured  &  Central & \multicolumn{5}{c|} {Systematic uncertainty (by source) [MeV]}   \\
   parameter   & value [MeV]  & stat &  reconstruction  &  unfolding  &  acceptance  &  combined \\
 \hline
  $m_{\mathrm 3h}^{\mathrm{cut}}$  &  591  & $\pm$  2 &   $\pm$ 6 & $\pm$ 4 &  $-10$ &    ${+7.5}/{-13}$\\  
\hline
  $Q_{\mathrm LS}$  & ~~89.7 &  \multicolumn{3}{c|}{ $\pm$ 2.1 }   &  $-2.8$  &  ${+2.1}/{-3.3 }$ \\
$\sigma_{\mathrm LS}$   & ~~44.3  &     \multicolumn{3}{c|}{ $\pm$ 0.8 }   &  $-1.0$ &  ${+0.8}/{-1.3 }$ \\
  $Q_{\mathrm OS}$  & 256.4 & \multicolumn{3}{c|}{ $\pm$ 5.5 }   &  $-7.3$ & ${+5.5}/{-9.1 }$\\
 $\sigma_{\mathrm OS}$   & ~~44.2 &   \multicolumn{3}{c|}{ $\pm$ 1.9 }   &  $-2.6$ & ${+1.9}/{-3.2 }$\\
\hline 
\end{tabular}
\end{center}}
\end{table}

   The variation of the acceptance requirements is also used to examine the variation of the correlation strength
 with charged-particle multiplicity in the inclusive sample, independently of the chain selection.
  Table~\ref{table:rate} summarizes the results. The systematic uncertainty is a combination of the uncertainty of the folding factors (5\%)
 and the bias due to the non-zero total charge in a reconstructed event (15\%), the latter being evaluated with help of 
 subsample of events with balanced content of positively- and negatively-charged tracks.
 The correlation strength remains stable within the restricted pseudorapidity region, despite the sharp
 drop of the average charged-particle multiplicity (factor 0.33). Such a behavior supports the hypothesis of a
 linear dependence of correlations on the number of particles.
 The measured correlation strength corresponds to the minimum rate of three-hadron chains per charged particle
 required to reproduce the correlation shape of the inclusive $\Delta(Q)$.
 The observed correlation strength in the baseline acceptance region is rather low, $1.1^{+0.1}_{-0.2}$\%, as reported in
  Table~\ref{table:rate}.
 This suggests that it is sufficient to have a single three-pion chain in the ground-state per three events to generate enough correlations
 to reproduce the data.

  The correlation strength is reduced by a factor of 2 with the increase of the transverse momentum threshold to 200~MeV.
 In the minimum-bias sample, strings are oriented predominantly along the beam axis, and therefore
 the observed variation of the correlation strength suggests the correlated hadrons have
 small intrinsic $p_{\mathrm T}$. The quantized fragmentation model predicts the correlations occur between pions with
 intrinsic transverse momentum of $\sim$ 134~MeV, with respect to the string axis. The detailed verification
 of the low-$p_{\mathrm T}$ dependence of the chain properties is a challenging task for future studies.

\begin{table}[h!]{
\caption{ Variation of the charged-particle multiplicity and of the
  correlation strength with the change of acceptance region. $N_{\mathrm ch}^{\mathrm main}$ stands for the corrected charged-particle multiplicity in the acceptance region of the base analysis. All numbers are corrected for detector effects.\label{table:rate}}
\begin{center}
\begin{tabular}{|l|c|c|c|}
\hline
  Acceptance   &  $p_{\mathrm T}>$100 MeV &  $p_{\mathrm T}>$100 MeV & $p_{\mathrm T}>$200 MeV      \\
   variations   & $|\eta| <$ 2.5  &  $|\eta| <$ 1  & $|\eta| <$ 2.5       \\
\hline
  $N_{\mathrm ch}/N_{\mathrm ch}^{\mathrm main}$ & 1 (by construction) &  0.33 & 0.78   \\
  $-\int_{\Delta Q<0}d\Delta Q$ [\%] &  1.07$\pm$0.03(stat)$^{+0.05}_{-0.17}$(syst) & 1.24$\pm$0.07(stat)$^{+0.06}_{-0.21}$(syst) & 
  0.56$\pm$0.03(stat)$^{+0.03}_{-0.10}$(syst) \\
\hline 
\end{tabular}
\end{center}}
\end{table}

\subsection{ Comparison with MC models }

\begin{figure}[htb!]
\subfigure[ ]{
  \label{delta_mc_incl}
\includegraphics[width=0.48\textwidth]{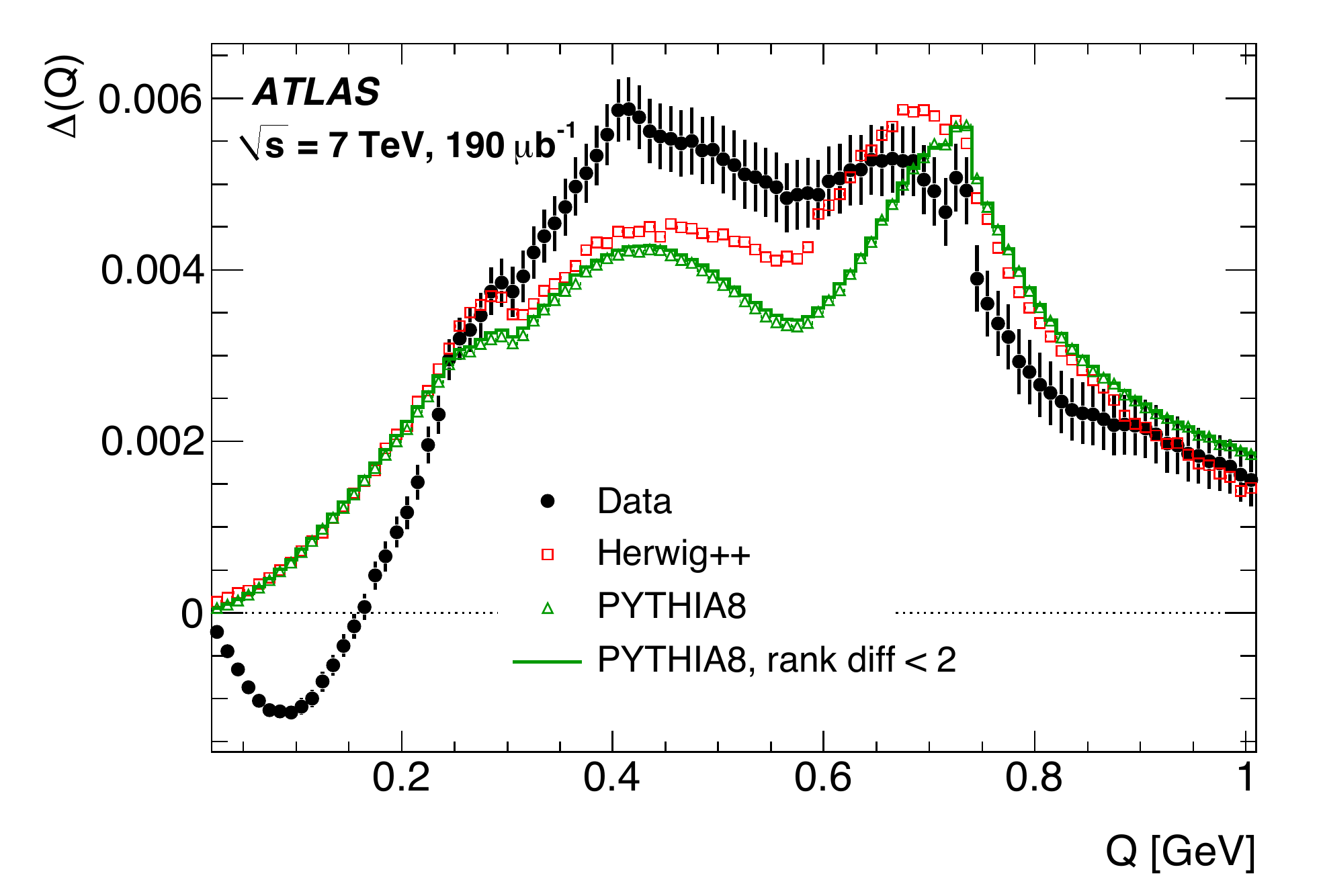}
}
\subfigure[ ]{
  \label{delta3h_mc}
\includegraphics[width=0.48\textwidth]{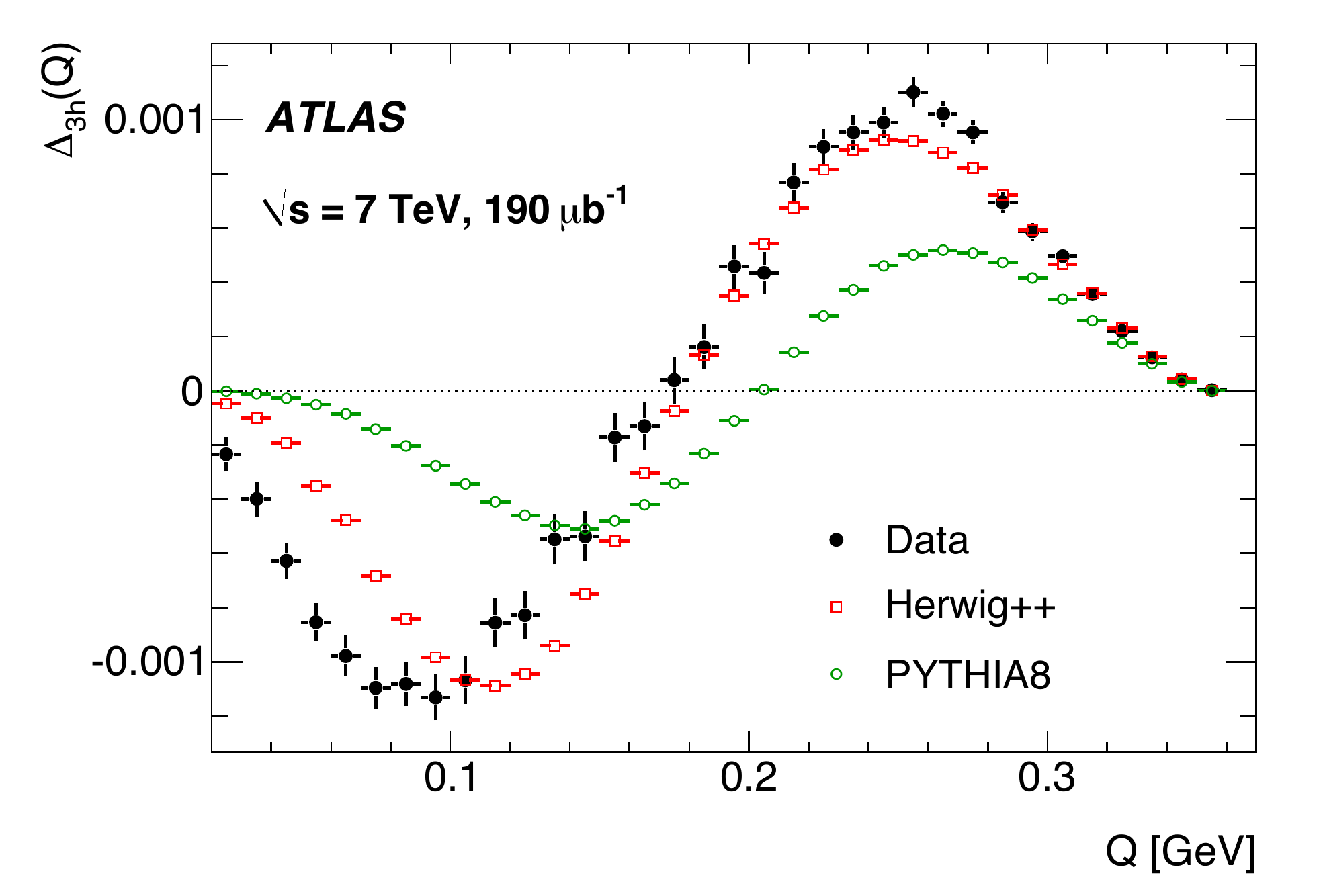}
}
\caption{ 
 (a) The comparison between the measured inclusive $\Delta(Q)$ and the prediction of the  Lund string fragmentation model
(represented by PYTHIA) and of the cluster hadronization model
(represented by Herwig++). In the Lund string fragmentation model, the
inclusive $\Delta$ distribution reflects the distribution of true adjacent
hadrons pairs (rank difference<2), indicated by the full line. 
 (b) Comparison of the measured $\Delta_{\mathrm 3h}(Q)$ correlation function with a representative set of hadronization models,
 for the three-hadron chain selection with mass $m_{\mathrm 3h}<$ 0.59~GeV.
\label{fig:mc}}
\end{figure}

 The comparison of the unfolded $\Delta(Q)$ distribution with the predictions
of PYTHIA and Herwig++ is shown in Figure~\ref{delta_mc_incl}. The Lund fragmentation model and
the cluster hadronization model predict a very similar distribution in the region of interest for the current
analysis ($Q \leq$ 0.3) and both disagree with the data significantly. There is a close correspondence
between the inclusive $\Delta(Q)$  observable and the distribution of
the momentum difference between true adjacent pairs produced by the string fragmentation, retrieved from the PYTHIA8 event record. 
This is a feature discussed in Section~\ref{correlation_function}: the inclusive $\Delta(Q)$ is constructed in a way which
makes is suitable for study of adjacent hadron pairs. 
Since the integral of $\Delta(Q)$ is an invariant, there is a direct relation between the excess of 
measured hadron pairs in the region of $0.3 < Q < 0.6 \ $~GeV, with respect to the MC models, and the
depletion at $Q<0.2$~GeV. The hypothesis of a physics threshold in
the emission of adjacent hadrons offers a plausible explanation of the discrepancy between
the data and the models. According to the MC estimates, the heavy flavor contribution to the low-$Q$ spectrum
is negligible.
 
Figure~\ref{delta3h_mc} shows the comparison of the measured $\Delta_{\mathrm 3h}$ with PYTHIA and Herwig++ predictions.
PYTHIA does not describe the data and the study of the event record reveals that the shape of its prediction is
dominated by chains formed by a pair of adjacent hadrons (with opposite charges) and a hadron originating
from another string. Herwig++ describes the chain selection much better even though it fails to reproduce the low-$Q$
part of the measured distribution. A similar observation was made in
the study of the azimuthal ordering~\cite{STDM-2011-20}, where some observables were better described by Herwig++.

Given that neither of these event generators contains the correlation effects induced by the fragmentation of the helical QCD string,
yet the data are to some extent described, the presence of chains of ground state hadrons cannot be assessed
from the shape of the measured $\Delta_{\mathrm 3h}$ alone.  
A further study of the properties of the selected hadron triplets is discussed in the next section.
  
\subsection{ Three-body decay }  \label{three-body}
 The selected three-hadron chains with mass below 0.59~GeV 
 are used to fill a Dalitz plot with coordinates ($X$,$Y$) defined in Eq.~(\ref{eqn:dalitz}).
 The measurement has the advantage of
 providing a single entry for each selected hadron triplet (instead
 of three separate entries in $\Delta_{\mathrm 3h}$) and therefore provides
 more direct information than the measurement of the correlation shape. However,
 the combinatorial background is no longer subtracted, which means the
 correlations are studied on top of a large background distribution.

\begin{figure}[ t!]
\begin{center}
\includegraphics[width=0.48\textwidth]{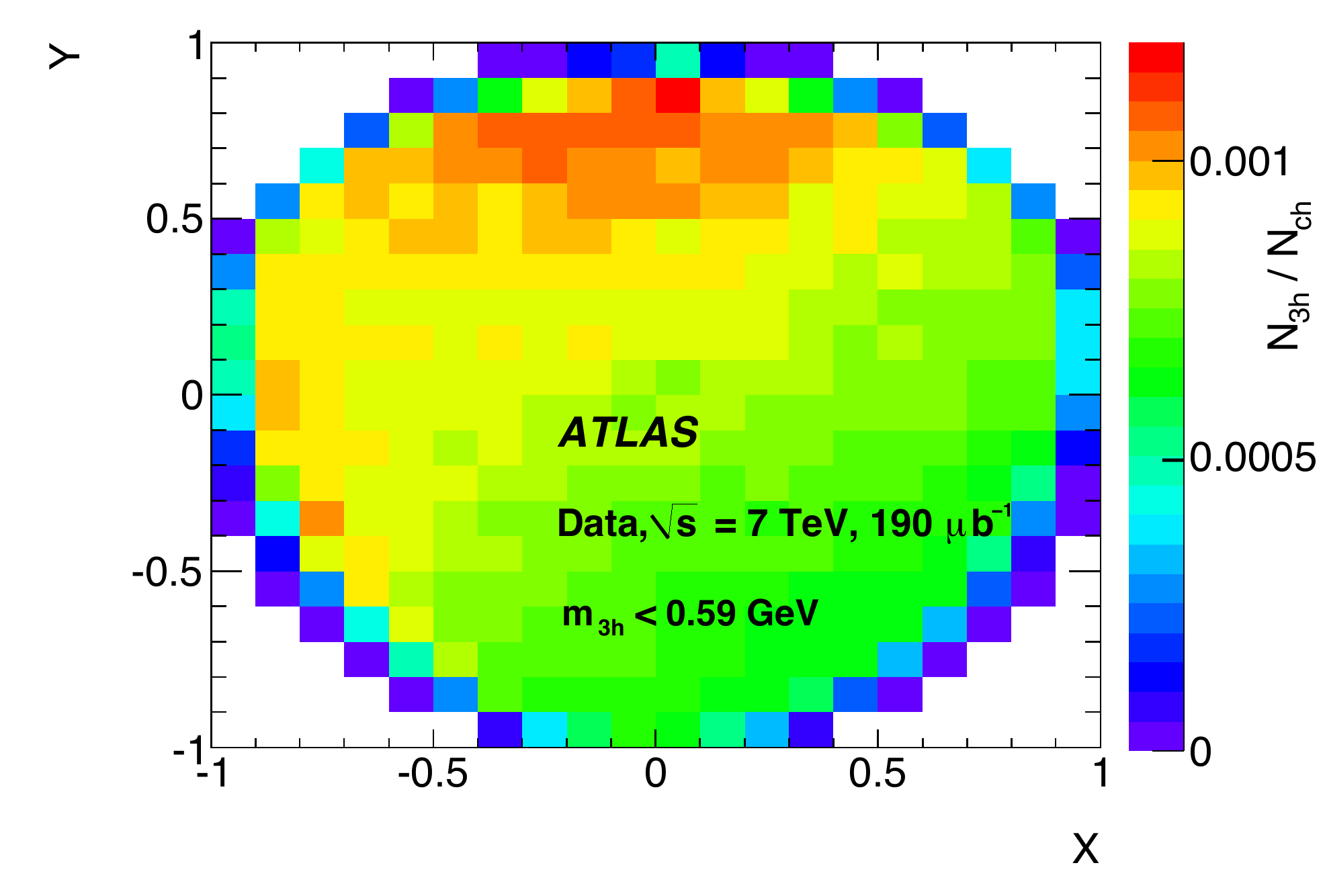}
\includegraphics[width=0.48\textwidth]{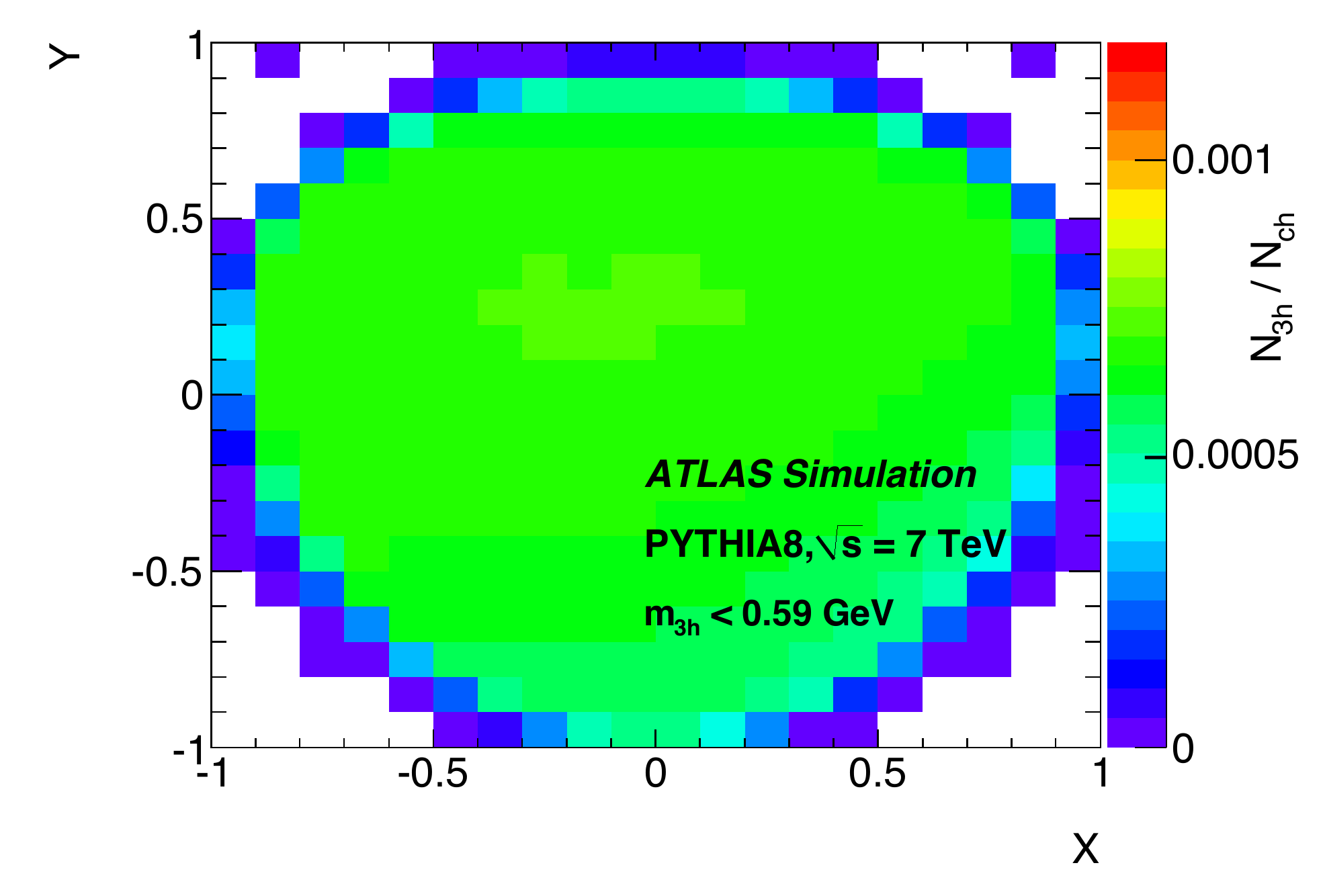}
\includegraphics[width=0.48\textwidth]{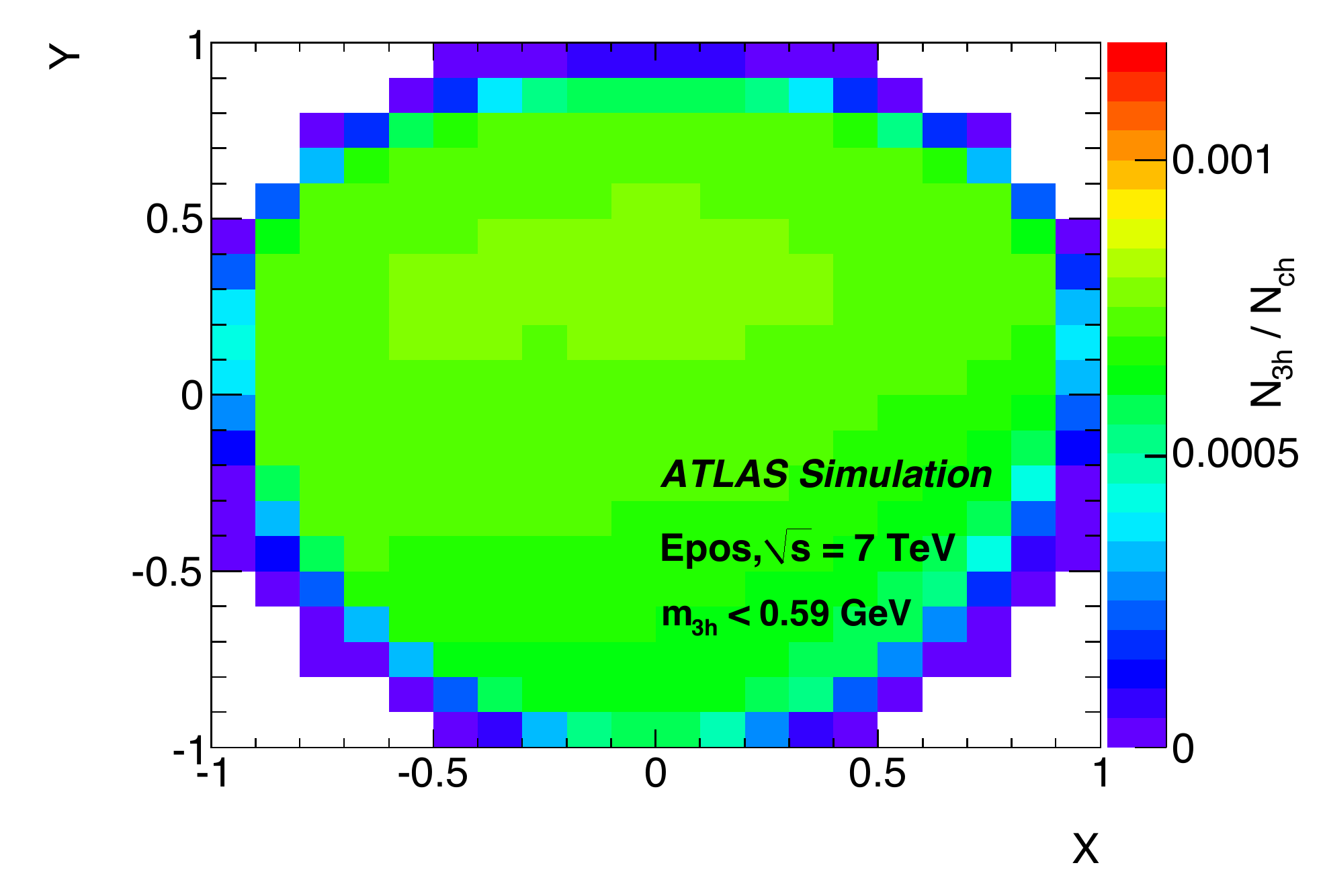}
\includegraphics[width=0.48\textwidth]{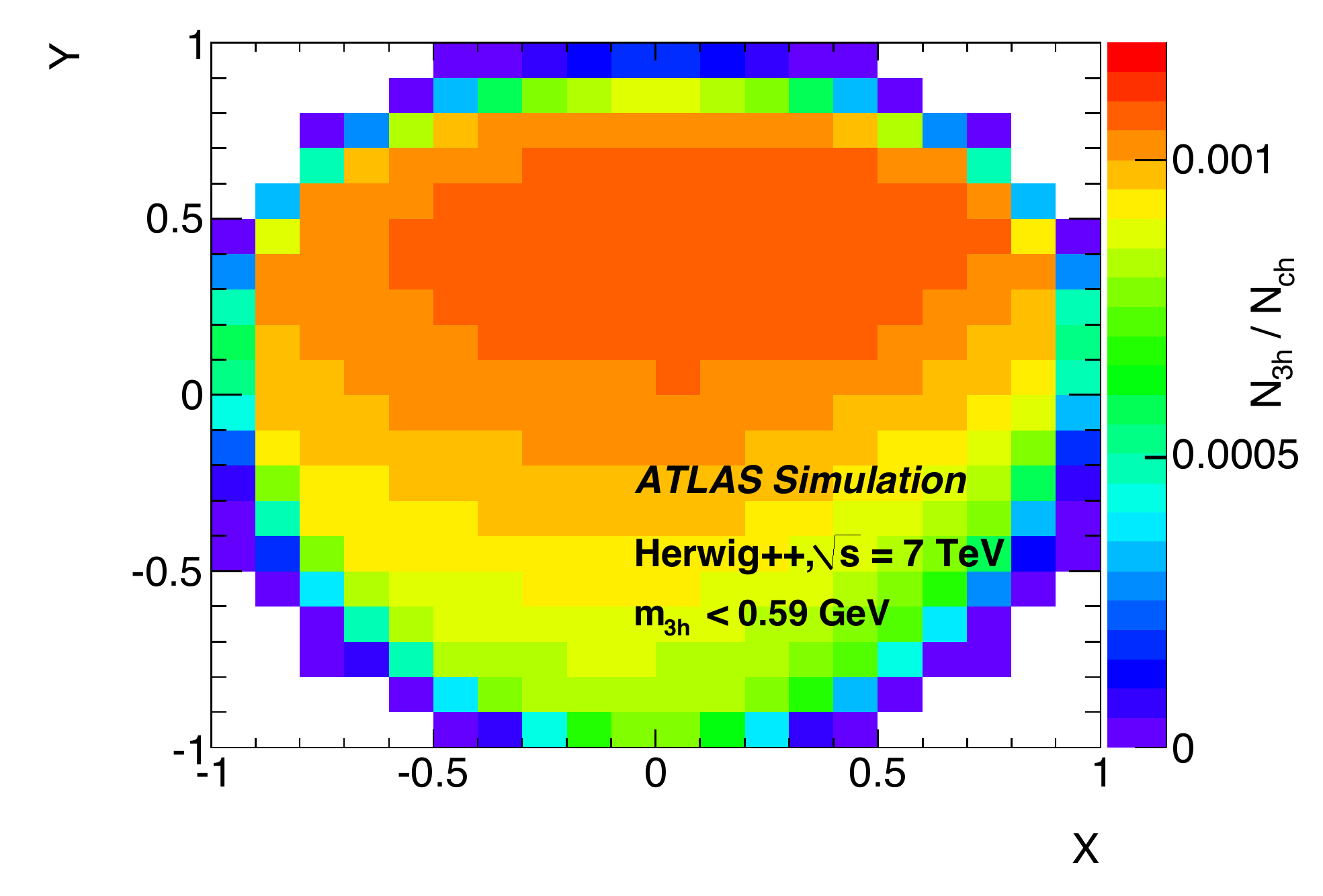}
\caption{ The Dalitz plot, Eq.~(\ref{eqn:dalitz}), filled with the three-body decay pattern of the three-hadron chain
selection ($m_{\mathrm 3h}<$~0.59~GeV). Top left: ATLAS data with detector effects
unfolded. Top right: PYTHIA8 4C prediction for minimum-bias events.
Bottom left: Epos prediction for minimum-bias events. Bottom right: Herwig++ prediction for minimum-bias events.
\label{fig:dalitz}}
\end{center}
\end{figure}

 The data are corrected with help of the HBOM estimate of the chain-recombination
 and of the track-based weight factors as described in Section~\ref{correction}. The correction procedure was verified with the
 PYTHIA6 sample. 
 Figure~\ref{fig:dalitz} shows the corrected decay pattern of selected
 chains in the data and the equivalent obtained from the generator-level MC samples. The distributions are normalized to the number
 of charged particles in the sample.

  The rate of selected chains is $0.24 \pm 0.02$ per charged particle in the
 data, 0.18 in PYTHIA8, 0.20 in Epos and 0.28 in Herwig++.   A clear
  enhancement is observed in the data at large $Y$ values, which is not reproduced by any of the MC samples
  (the quantum effects affecting the hadron momenta are not included in the event generators). 

 \begin{figure}[t!]
\begin{center}
\includegraphics[width=0.49\textwidth]{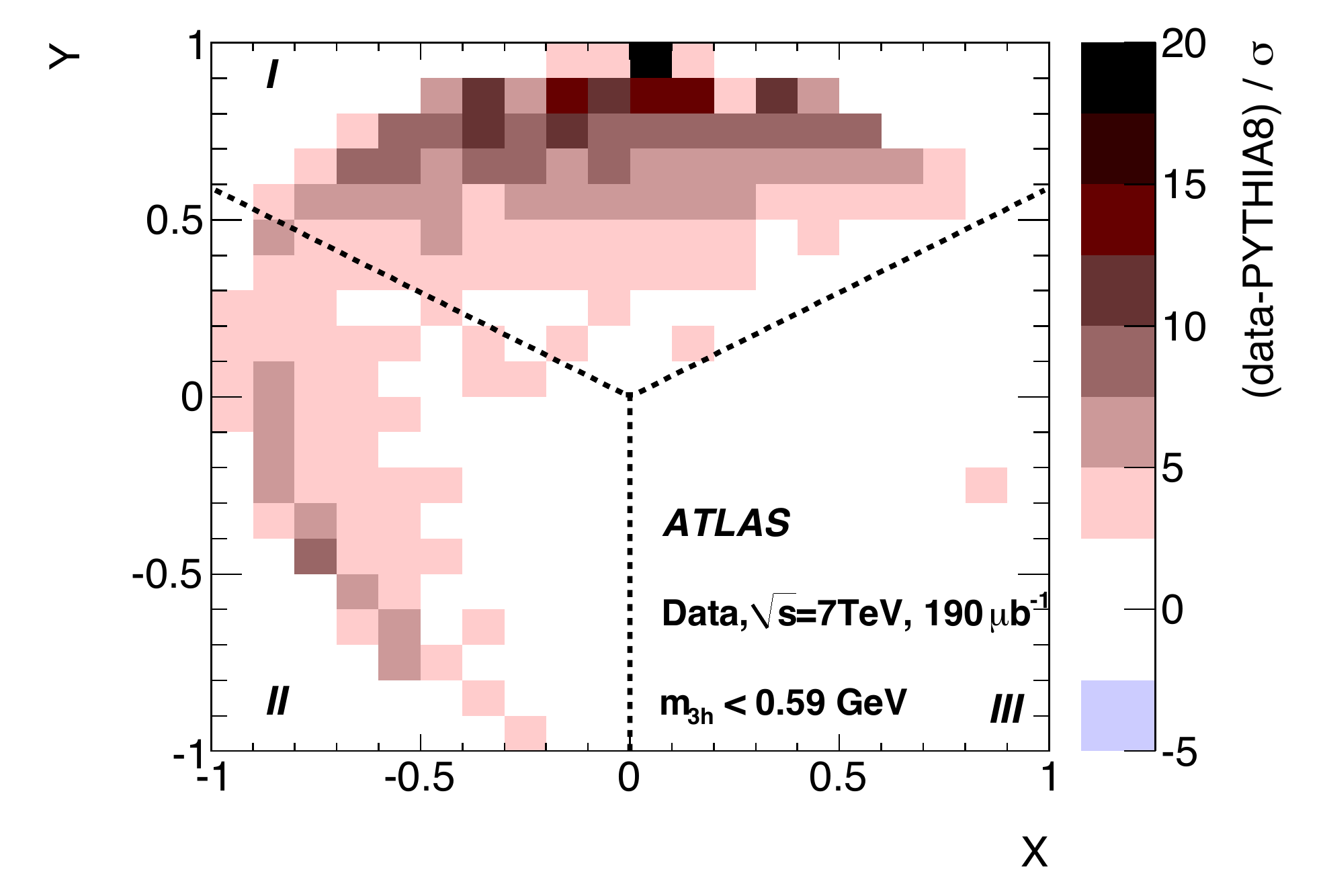}
\includegraphics[width=0.49\textwidth]{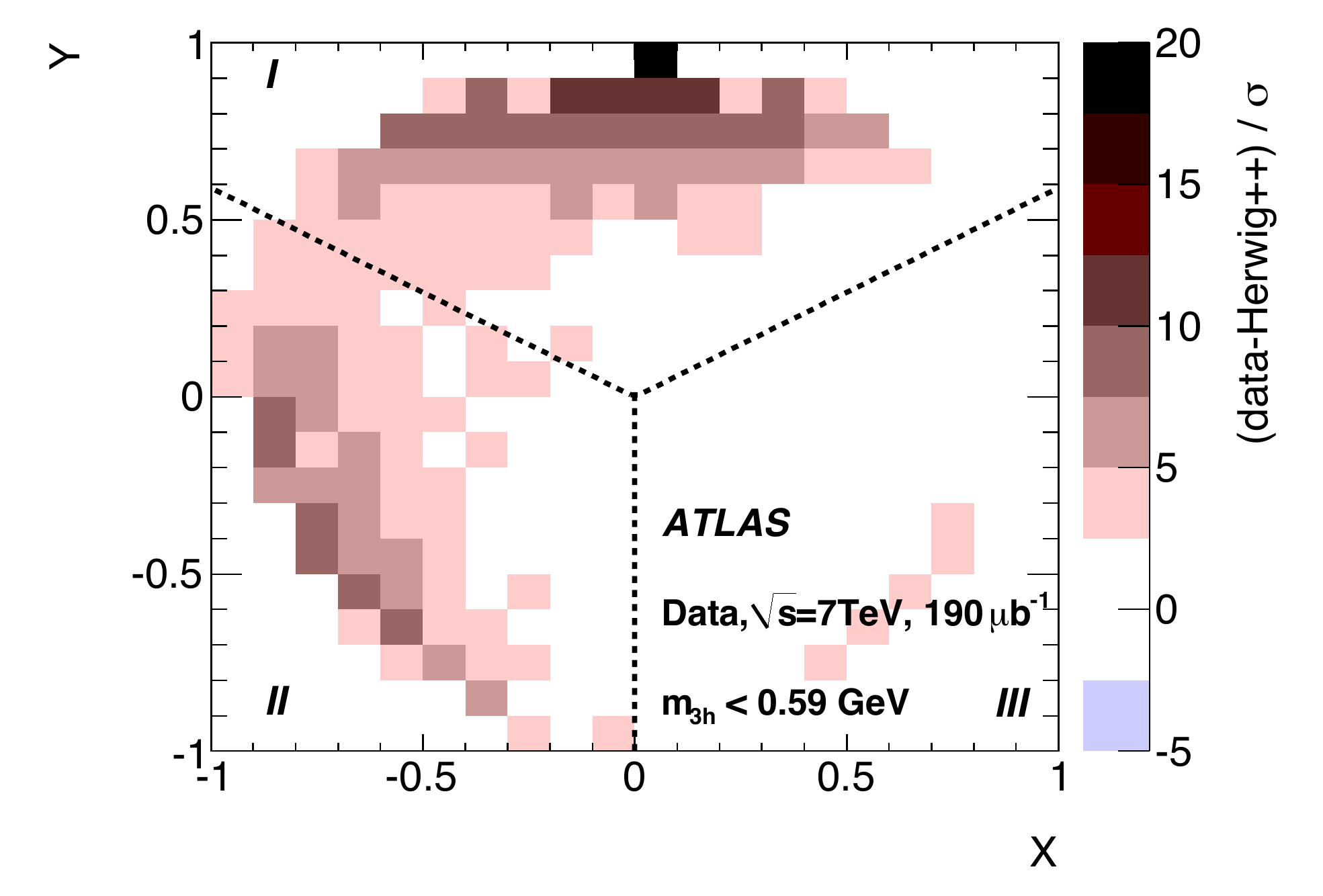}
\caption{ The significance of the difference of the Dalitz
  plot filled with the three-body decay pattern of chains with mass below
  0.59~GeV, between the data and the PYTHIA8 simulation (left) and the
  Herwig++ simulation (right). 
  The signal region I gathers chains where
  the pair of like-sign hadrons carries the least momentum difference. Region III serves as a reference for the
  adjustment of the MC simulation with respect to the data. 
\label{fig:significance}}
\end{center}
\end{figure}     
 
  In order to quantify the significance of the shape difference
  between the data and the MC models, the Dalitz plot is split into three regions, see Figure~\ref{fig:significance}. The principal signal
  region I gathers chains where the pair of like-sign hadrons carries the smallest momentum difference. A less
  pronounced but still quite significant excess is observed in region II
  where it can be interpreted as a signature of a quadruplet chain
  containing a neutral pion (see below). The models 
  are rescaled to reproduce the rate of selected chains in the reference region III where no
  enhancement is observed and where no enhancement is expected by the
  quantized helix string fragmentation model (see Section~\ref{interpretation}).
  The significance of the difference  between the data and the MC prediction is then evaluated as the number of
  standard deviations observed, calculated from the statistical error and from the
  uncertainty associated with the bin migration and with the $X,Y$ dependence of track-based weights and of the presence of non-primary tracks.
  The reconstruction uncertainty
  as well as the correlated uncertainty of folding/weighting factors is strongly correlated between
  bins and therefore absorbed in the adjustment of the MC chain rate. 
  The $\chi^2$ per degree of freedom  for comparison of data with PYTHIA8  is
  29.1, 8.1 and 0.89 in regions I, II and III 
  respectively.  The comparison of data with Herwig++ yields similar
  results: $\chi^2$ per degree of freedom  = 20.0, 12.4 and 0.98 in regions I, II
  and III respectively. 
  The excess in region I, which
  contains triplets where the like-sign pair of particles
  carries the smallest momentum difference is $1.72 \pm 0.05$\% ($1.35 \pm 0.05$\%)  of chains per charged particle compared to the adjusted PYTHIA8(Herwig++) shape.
  The excess in region II is $0.75 \pm 0.04$\% ($0.97 \pm 0.04$\%) of chains per charged particle compared to the PYTHIA8(Herwig++) shape.

\section{Interpretation of measurements using the helix string model} \label{interpretation}  

  The observations described in the previous section agree with the hypothesis that the observed correlations 
  reflect the pattern of coherent emission of hadrons from a helical QCD
  string. Since the adjusted chain mass limit agrees well with the expected
  minimum mass for a true chain of charged pions,
 the position of the maximum of the $\Delta_{\mathrm 3h}(Q)$ distribution can be interpreted as the measurement of the momentum
 difference between adjacent opposite-sign pairs of pions with rank
 difference 1, while the position of the minimum can be associated with the 
 momentum difference between like-sign pairs of pions with rank difference~2. 

\subsection{Onset of adjacent hadron pair production} \label{threshold}

  The existence of a quantum threshold for the minimum momentum difference between adjacent hadrons is a fundamental
feature of the quantized fragmentation model. The data are in agreement with the prediction of a threshold-like behavior:
  after the subtraction of selected three-hadron chains from the inclusive $\Delta(Q)$,
 there are no adjacent pairs visible in the low Q region, up to a
 certain threshold value, which depends on the assumption made about
 the length of correlated hadron chains:
\begin{eqnarray}
    \Delta^{\mathrm A}(Q) & = & \Delta(Q) - \Delta_{\mathrm 3h}(Q),  \label{eqn:deltaA} \\
    \Delta^{\mathrm B}(Q) & = & \Delta(Q) - \Delta_{\mathrm 3h}(Q) - f_{\mathrm OS}(Q;Q_{\mathrm OS},\sigma_{\mathrm OS}).   \label{eqn:deltaB}
\end{eqnarray}    
  Hypotheses A and B refer to Figure~\ref{fig:chainflow}. Hypothesis A describes the chain contribution as a contribution from 
  long uninterrupted chains. Hypothesis B assumes the triplet chains are disconnected and restores the contribution from opposite-sign pairs,
  which had been scaled by factor 0.5 in Eq.~(\ref{eqn:delta3h}), using the fitted decomposition of $\Delta_{\mathrm 3h}$ into two Gaussian functions, Eq.~(\ref{eqn:gaus}). 

\begin{figure}[htb!]
\begin{center}
\includegraphics[width=0.6\textwidth]{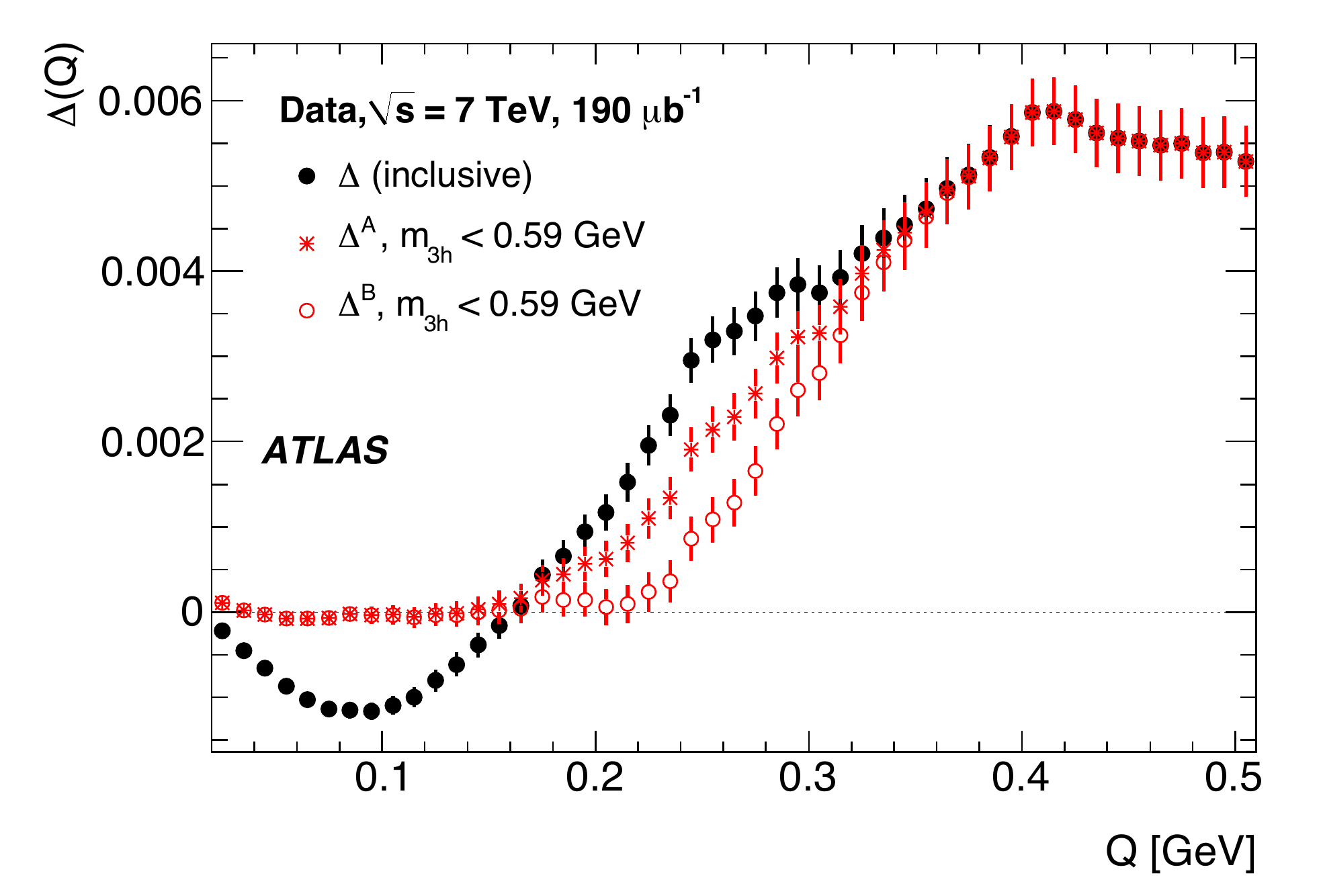}
\caption{ 
  The unfolded $\Delta(Q)$ with (black points)  and without (red points) the contribution from low-mass three-hadron
 chains. Closed squares indicate the subtraction performed assuming long uninterrupted chains (A), open circles
 indicate the subtraction done assuming disconnected triplets (B).   
\label{fig:threshold}}
\end{center}
\end{figure}

 The distributions after the subtractions are shown in Figure~\ref{fig:threshold}. In the scenario of disconnected three-pion chains, the threshold value moves up to $\sim$ 0.25~GeV. This value coincides
 with the quantum threshold predicted by the quantized model of fragmentation of a helical QCD string, 
 which also fits the experimentally found position of the peak formed by closest opposite-sign pairs.

\subsection{Enhanced production of pairs of like-sign charged hadrons} \label{be}

  The enhanced production of pairs of like-sign charged hadrons is
  traditionally attributed to the Bose--Einstein effect, originating in the
 symmetrization of the quantum-mechanical amplitude
 with respect to the exchange of identical bosons~\cite{STDM-2013-01}.
 A large number of measurements have been done on the basis of the correlation function defined as a ratio
 of like-sign and opposite-sign distributions, the latter being considered ``uncorrelated'' by the Bose--Einstein formalism.

\begin{figure}[h!]
\begin{center}
\includegraphics[width=0.6\textwidth]{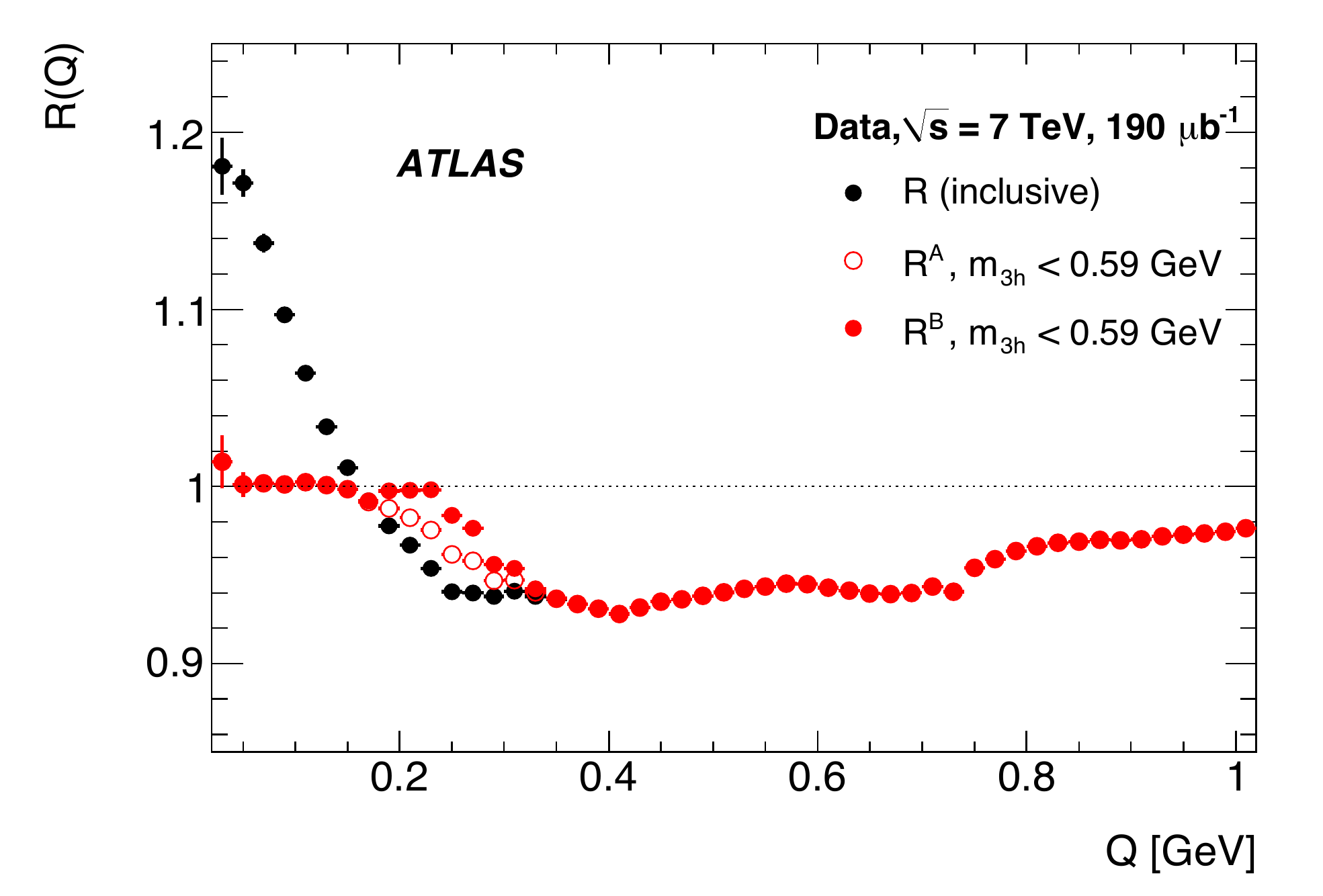}
\caption{ 
 The ratio of inclusive like-sign and opposite-sign pair spectra, before and after subtraction
 of the estimated contribution from low-mass three-hadron chains, in
 scenario A (long uninterrupted chain)  and B (disconnected triplet
 chains). The upper mass limit for the three-hadron chain (0.59~GeV)
has been set to the value that reproduces the enhancement of like-sign pair production in the inclusive sample.
\label{fig:ratio}}
\end{center}
\end{figure}

  Figure~\ref{fig:ratio} shows the ratio $R=N(Q)^{\mathrm LS}/N(Q)^{\mathrm OS}$ 
 before and after subtraction of the estimated contribution from ordered hadron chains with mass below 0.59~GeV. The subtraction is done
 in close analogy to Eqs.~(\ref{eqn:deltaA}) and (\ref{eqn:deltaB}) with the help of the fit (Eq.~(\ref{eqn:gaus})):
\begin{eqnarray}
    R^{\mathrm A}(Q) & = & [N(Q)^{\mathrm LS} + N_{\mathrm ch} \ f_{\mathrm LS}(Q;Q_{\mathrm LS},\sigma_{\mathrm LS})]/ [N(Q)^{\mathrm OS} - N_{\mathrm ch} \ f_{\mathrm OS}(Q;Q_{\mathrm OS},\sigma_{\mathrm OS})],   \\
    R^{\mathrm B}(Q) & = & [N(Q)^{\mathrm LS} + N_{\mathrm ch} \ f_{\mathrm LS}(Q;Q_{\mathrm LS},\sigma_{\mathrm LS})]/ [N(Q)^{\mathrm OS} - 2 N_{\mathrm ch} \ f_{\mathrm OS}(Q;Q_{\mathrm OS},\sigma_{\mathrm OS})],  
\end{eqnarray}    
  i.e. the $\Delta_{\mathrm 3h}$ distribution is decomposed into contributions from like-sign and opposite-sign pairs, which are subtracted from their respective inclusive two-particle distributions, before calculating the ratio.    
 The subtraction is done for the hypothesis of long chains [A](open red points) and disconnected chains [B](closed red points).
 In both cases, the chain selection
 contains the source of the enhanced like-sign pair production, hence providing an alternative explanation of the data. 

\subsection{ Three-body decay properties of quantized fragmentation} 

 \begin{figure}[hbt!]
\begin{center}
\includegraphics[width=0.49\textwidth]{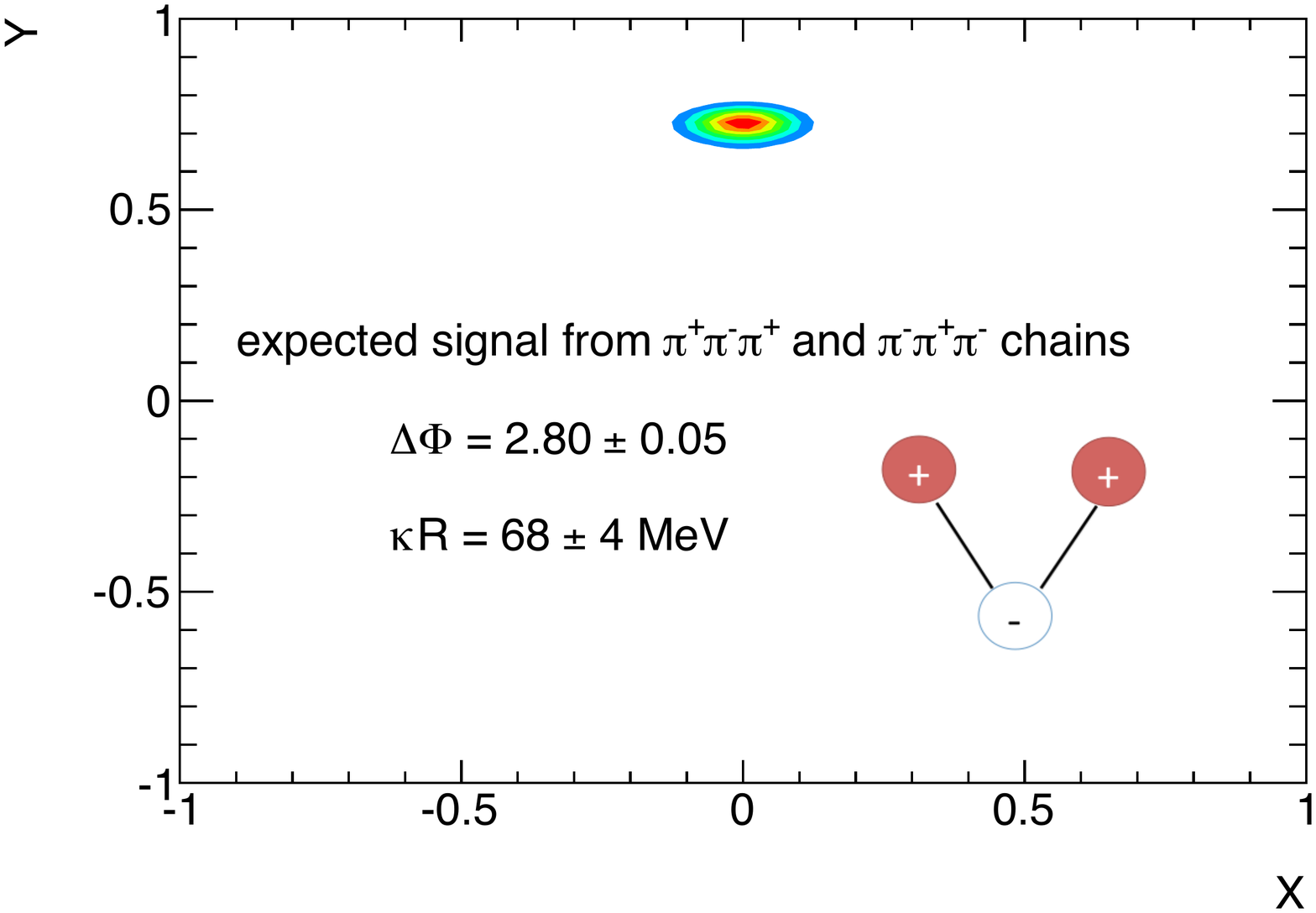} 
\includegraphics[width=0.49\textwidth]{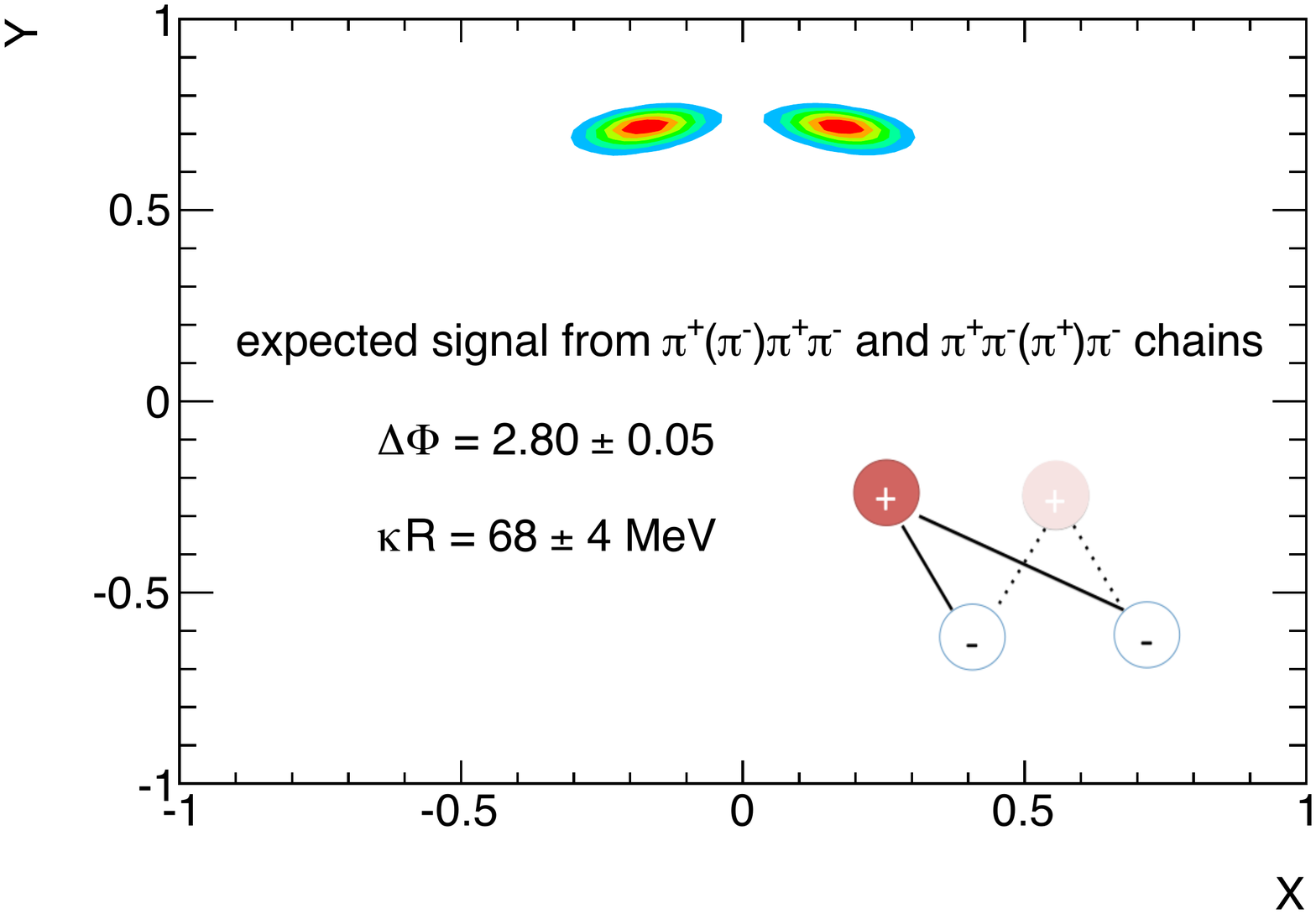} 
\includegraphics[width=0.49\textwidth]{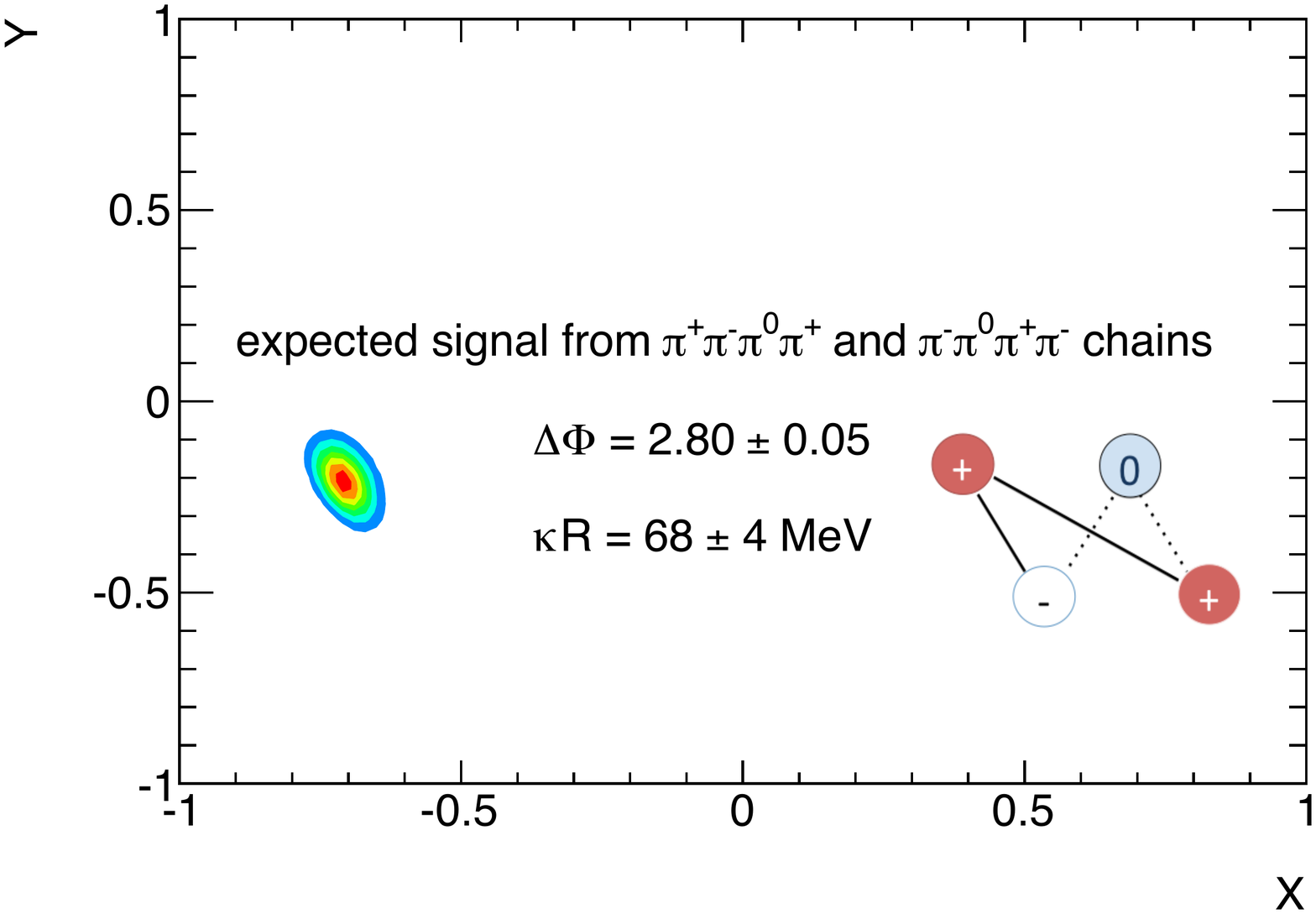} 
\includegraphics[width=0.49\textwidth]{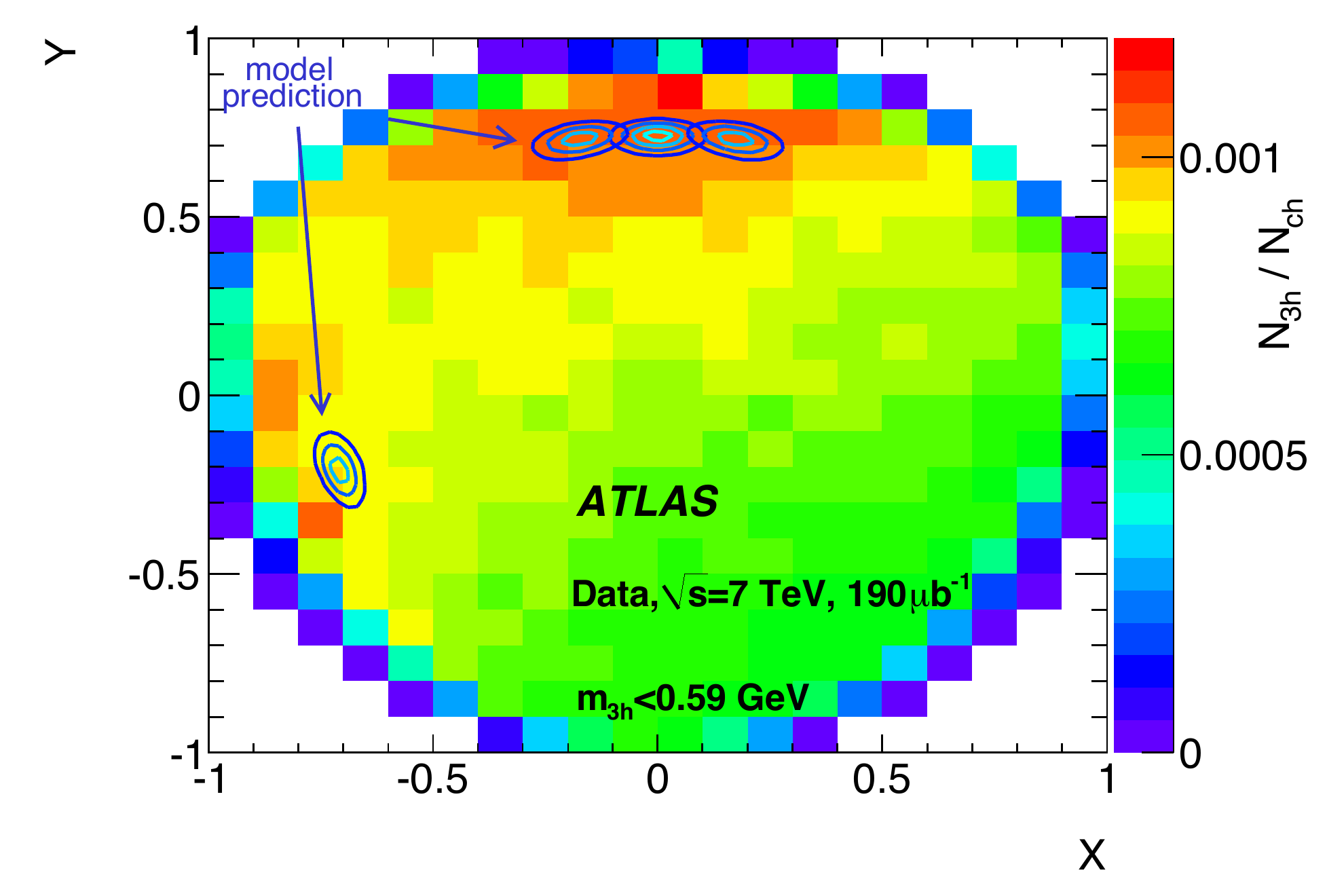} 
\caption{ Model prediction for the decay pattern of
  ground-state triplet (top left), ground-state quadruplet with missing
  middle member (top right) and ground-state quadruplet containing a
  neutral pion (bottom left), in arbitrary z scale. Model predictions are calculated 
  using a Gaussian smearing of helical string parameters constrained by the fit of mass spectrum of 
  pseudoscalar mesons. The
  resolution uncertainty of the order of 10~MeV is included in the
  variation of string parameters. 
  Bottom right: Comparison of observed chain decay pattern with model predictions.
\label{fig:dalitz_exp}}
\end{center}
\end{figure}

 Figure~\ref{fig:dalitz_exp} compares the measured three-body decay pattern
 of the chain selection with the model prediction obtained by generating hadron
 triplets and quadruplets according to the pattern of helical string fragmentation
 into chains of ground-state pions (see Appendix~\ref{sec:appendix}). String parameters
 $\Delta\Phi $ and $\kappa R$ are smeared by 2\% and 6\%, respectively.   The maximal
 enhancement in the data corresponds to the expected location of the
 signal from the chain of three ground-state pions, with some
 indication for a possible presence of quadruplet chains:  
 the wide ``shoulder''  of the signal in region I can be interpreted
  as a signature   of a quadruplet chain with a missing
  middle particle.
  The observation of such a form of signal has implications
  for the interpretation of the results of the measurements.  If there is a significant admixture of opposite-sign pairs with
 rank difference 3 (for which the momentum difference should be around 0.236~GeV),
 the bias should be taken into account in the interpretation of
 results in terms of properties of the QCD string.
 This bias is obtained from the best fit of the Dalitz plot with the mixture of PYTHIA6 shape (for background) and from the model predictions
 for the triplet and for the incomplete-quadruplet chains (for the signal) and found to be  $-9 \pm 5$~MeV. The uncertainty is estimated 
  as one third of the case, where the selection is equally spread between pairs with rank difference 1 and 3.

 As described in the introduction, the quantized helical string fragmentation model draws input from the mass spectrum of light
 pseudoscalar mesons. It is therefore possible that their decay
 contributes to the visible correlation pattern, but such a
 measurement would require the reconstruction of $\pi^0$ and therefore falls
 beyond the scope of the current analysis. 

\subsection{ Momentum difference as a function of hadron rank difference } \label{measured_helix}

    Taking into account the uncertainty of the chain selection (Table~\ref{tab:msys}), the
    inclusive two-particle correlation pattern is reproduced by 
   three-hadron chains below a mass limit of 
   \begin{equation}
      m_{\mathrm 3h}^{\mathrm{cut}} = 591 \ \pm \ 2  \ {\mathrm(stat)} \ ^{ + \ 7.5}_{-13}  \ {\mathrm
        (syst)  \ MeV. }
\end{equation} 

     The data show a threshold effect in the production
  of adjacent hadron pairs. The threshold coincides with the emergence of the Gaussian peak situated at

  \begin{equation}
      Q_{\mathrm OS} = 256.4 \pm 0.5 \  {\mathrm (stat) } \  \pm 1.8 \ {\mathrm (rec)}
     \  ^{+5.5}_{-9.1}   \ {\mathrm   (chain \ selection) }  \  {\mathrm MeV},
   \end{equation}

   which is obtained from the fit of the preferred momentum difference
   between opposite-sign pairs in the selected chains.  Taking
   into account the possible admixture of pairs with rank difference 3, suggested by the
   particular form of the three-body decay pattern of selected hadron
   chains, the best estimate of the momentum difference for hadron
   pairs of rank difference $r$ = 1 becomes

  \begin{equation}
      Q(r=1) = 265.6 \pm 0.5 \  {\mathrm (stat) } \  \pm 1.8 \ {\mathrm (rec)}
     \  ^{+7.4}_{-10}   \ {\mathrm   (chain \ selection) }  \  {\mathrm MeV} .\label{eqn:qr1}
   \end{equation}

   The preferred momentum difference
   between hadrons with like-sign charge combination (rank difference r = 2) is found to be

  \begin{equation}
      Q(r=2) = 89.7 \pm 0.4 \  {\mathrm (stat) }  \pm 1.2 \ {\mathrm
        (rec)} \  ^{+2.1}_{-3.3}  {\mathrm (chain \ selection) } \  {\mathrm
        MeV}  \label{eqn:qr2}
   \end{equation}

 for the same set of selected three-hadron chains. Both like-sign and opposite-sign pair distributions have
 a Gaussian shape with a width of $44 \pm 3$~MeV, while the experimental resolution in the fitted region is better than 10~MeV.
 The systematic uncertainties in Eqs.~(\ref{eqn:qr1}) and~(\ref{eqn:qr2}) are correlated.    

   These values are in good agreement with the predictions of
 the model of a QCD string with a helical shape, with parameters constrained by mass spectrum of pseudoscalar mesons (see Table~\ref{tab:qrank}).  
   
\section{Conclusions} \label{sec:conclusions}       
    
   Two-particle correlation spectra measured in the minimum-bias sample at center-of-mass energy of 7 TeV are analyzed in the context of 
   coherent particle production. The data sample consists of 190 $\mu{\mathrm b}^{-1}$ of events produced with
   low-luminosity proton--proton beams at LHC and collected in the early 2010 ATLAS data taking. The QCD string
   fragmentation scenario is used to introduce the notion of ordered hadron chains. Using the assumption of
   the local charge conservation in the string breakup, the correlation function is
   defined in a way suitable for study of correlations between pairs and triplets of adjacent hadrons.
   Because it is experimentally impossible to assess the exact rank ordering of particles, the rank ordering
   is replaced by the minimization of the mass of hadron chains. The analysis relies on the removal of the
   background of random combinations by means of subtraction of pairs with like-sign charge combination from
   pairs with opposite-sign charge combination. The analysis does not rely on predictions of conventional MC models
   which fail to describe the data.
  
     The results indicate that the enhanced like-sign pair production at low-$Q$, observed in the data and 
  traditionally attributed to the Bose-Einstein effect, can be entirely attributed to the presence of ordered three-hadron chains
  with mass below $591 ^{+8}_{-13}$~MeV, at a minimum rate of $1.1  ^{+0.1}_{-0.2}$\% per charged particle.
  A strong dependence of the size of the effect on the transverse momentum of tracks in the laboratory frame is observed.
      
     The shape of the three-hadron chain contribution to the inclusive $Q$ spectra agrees with the hypothesis that these chains are
 produced via coherent quantized fragmentation of a homogeneous QCD string with a helical structure.
 The measured momentum difference between hadrons within such a chain is  $89.7 ^{+2.5}_{-3.5}$~MeV for pairs with rank difference 2 and $266 ^{+8}_{-11}$~MeV for pairs of adjacent hadrons.
   The data support the prediction of a ``forbidden'' region for the
   production of adjacent (opposite-sign) hadron pairs at low $Q$. The
   threshold is situated at $\sim$ 0.25~GeV and agrees with the
   quantum threshold predicted by the helical string model.

\section*{Acknowledgements}


We thank CERN for the very successful operation of the LHC, as well as the
support staff from our institutions without whom ATLAS could not be
operated efficiently.

We acknowledge the support of ANPCyT, Argentina; YerPhI, Armenia; ARC, Australia; BMWFW and FWF, Austria; ANAS, Azerbaijan; SSTC, Belarus; CNPq and FAPESP, Brazil; NSERC, NRC and CFI, Canada; CERN; CONICYT, Chile; CAS, MOST and NSFC, China; COLCIENCIAS, Colombia; MSMT CR, MPO CR and VSC CR, Czech Republic; DNRF and DNSRC, Denmark; IN2P3-CNRS, CEA-DSM/IRFU, France; SRNSF, Georgia; BMBF, HGF, and MPG, Germany; GSRT, Greece; RGC, Hong Kong SAR, China; ISF, I-CORE and Benoziyo Center, Israel; INFN, Italy; MEXT and JSPS, Japan; CNRST, Morocco; NWO, Netherlands; RCN, Norway; MNiSW and NCN, Poland; FCT, Portugal; MNE/IFA, Romania; MES of Russia and NRC KI, Russian Federation; JINR; MESTD, Serbia; MSSR, Slovakia; ARRS and MIZ\v{S}, Slovenia; DST/NRF, South Africa; MINECO, Spain; SRC and Wallenberg Foundation, Sweden; SERI, SNSF and Cantons of Bern and Geneva, Switzerland; MOST, Taiwan; TAEK, Turkey; STFC, United Kingdom; DOE and NSF, United States of America. In addition, individual groups and members have received support from BCKDF, the Canada Council, CANARIE, CRC, Compute Canada, FQRNT, and the Ontario Innovation Trust, Canada; EPLANET, ERC, ERDF, FP7, Horizon 2020 and Marie Sk{\l}odowska-Curie Actions, European Union; Investissements d'Avenir Labex and Idex, ANR, R{\'e}gion Auvergne and Fondation Partager le Savoir, France; DFG and AvH Foundation, Germany; Herakleitos, Thales and Aristeia programmes co-financed by EU-ESF and the Greek NSRF; BSF, GIF and Minerva, Israel; BRF, Norway; CERCA Programme Generalitat de Catalunya, Generalitat Valenciana, Spain; the Royal Society and Leverhulme Trust, United Kingdom.

The crucial computing support from all WLCG partners is acknowledged gratefully, in particular from CERN, the ATLAS Tier-1 facilities at TRIUMF (Canada), NDGF (Denmark, Norway, Sweden), CC-IN2P3 (France), KIT/GridKA (Germany), INFN-CNAF (Italy), NL-T1 (Netherlands), PIC (Spain), ASGC (Taiwan), RAL (UK) and BNL (USA), the Tier-2 facilities worldwide and large non-WLCG resource providers. Major contributors of computing resources are listed in Ref.~\cite{ATL-GEN-PUB-2016-002}.

\clearpage
\appendix
\part*{Appendix}
\addcontentsline{toc}{part}{Appendix}
\section{Calculation of predictions within the helical QCD string model } \label{sec:appendix} 

  In the model described in Ref.\cite{helix2}, the causal constraint applied to the helical QCD field leads to a quantization pattern describing the mass spectrum of
 hadrons with mass below 1~GeV. In particular, the pseudoscalar mesons ($\pi,\eta,\eta'$) can be regarded as string pieces 
 fragmenting into $(n = 1,3,5)$ ground-state hadrons (pions), with
 transverse energy ($E_{\mathrm T}$) and momentum ($\vec{p_{\mathrm T}}$) of mesons defined by
 the helical string properties (Figure~\ref{fig:string}):
\begin{equation}
  E_{\mathrm T}({n})  =  \sqrt{m_{n}^2 + p_{\mathrm T}({n})^2} =  {n} \ \kappa \ R  \ \Delta\Phi, \label{eqn:et}
\end{equation}
 where  $R$ stands for the radius of the helix, $\kappa \sim$~1~GeV/fm is the string tension, $\Delta\Phi$ is the quantized helix phase
difference describing the shortest piece of string that can form a
hadron, and $m_{n}$ is the (quantized) meson mass spectrum. 

 The transverse momentum of a hadron stems from the integral of the string tension along the trajectory (in the transverse plane)
 of a quark traveling from one breakup vertex to another. The string tension is tangential to the trajectory of the quark:
\begin{equation}
 \vec{p}_{\mathrm T}(\Phi,\Phi + n\Delta\Phi ) = (p_{\mathrm T}^x,i p_{\mathrm T}^y) = \kappa R \int_{\Phi}^{\Phi+n\Delta\Phi}{ {\mathrm e}^{\mathrm i(\alpha+\frac{\pi}{2})} \ d\alpha } = 2 \kappa R \ \sin{\frac{n\Delta\Phi}{2}} \  {\mathrm e}^{{\mathrm i}(\Phi+\frac{n\Delta\Phi}{2}+\frac{\pi}{2})}.
\label{eqn:pt}
\end{equation}

The fit of the mass spectrum of pseudoscalar mesons indicates a rather narrow radius of the helical string
 ($\kappa R = 68 \pm 2$~MeV) and a quantized phase difference $\Delta\Phi = 2.82 \pm 0.06$. These values
 translate into the quantized ground-state transverse energy, $E_{\mathrm  T} ({n}=1) \simeq $ 192~MeV,
 and the transverse momentum of a ground state pion, $p_{\mathrm T}(n=1) $ $\simeq$ 134~MeV~\cite{helix2}. 

When a piece of helical string  fragments into a chain of ground-state pions,
the string shape is reflected in the momentum difference of the emitted hadrons.
 Neglecting the longitudinal momentum differences between adjacent
hadrons i.e. assuming local homogeneity of the fragmenting QCD field,  the four-momentum difference between hadrons within a
chain of ground-state pions can be written as function of their rank
difference $r$:
\begin{equation}
  Q(r) = \sqrt{ -(p_i - p_{i+r}) ^2} \  = \ 2 \  p_{\mathrm T}({n=1}) \ | \  \sin{(r \ \Delta\Phi/2)} \  |,
\label{eqn:q}
\end{equation}
 where $r \Delta \Phi$
  corresponds to the opening angle between pions in the transverse plane. Eq.~(\ref{eqn:q}) is used to fill Table~\ref{tab:qrank}. 

 It is straightforward
 to calculate the (transverse) momenta of pions in the chain defined by string breakup vertices $\Phi_{j+1} = \Phi_{j}+\Delta\Phi, \cdots , 
 \Phi_{j+n}=\Phi_{j}+ n \Delta\Phi$, and to study the properties of pion multiplets for various rank difference combinations.
  A closer look at Table~\ref{tab:qrank} suggests the minimization of the mass of selected three-hadron clusters does not ensure the selection
 of a triplet of adjacent pions in case of presence of a longer chain. The largest contamination presumably comes from an incomplete quadruplet
 chain with a missing (internal) pion, and from a quadruplet chain containing a neutral pion. Quadruplet chains are therefore included as a correction to the model predictions.      
  For the study of three-body decays, pions in selected triplets are boosted into the rest frame of the triplet, and their kinetic energy is evaluated
 (the longitudinal momenta of triplet members are negligible in the rest frame of the triplet). 

\begin{figure}[hbt!]
\begin{center}
\includegraphics[width=0.7\textwidth]{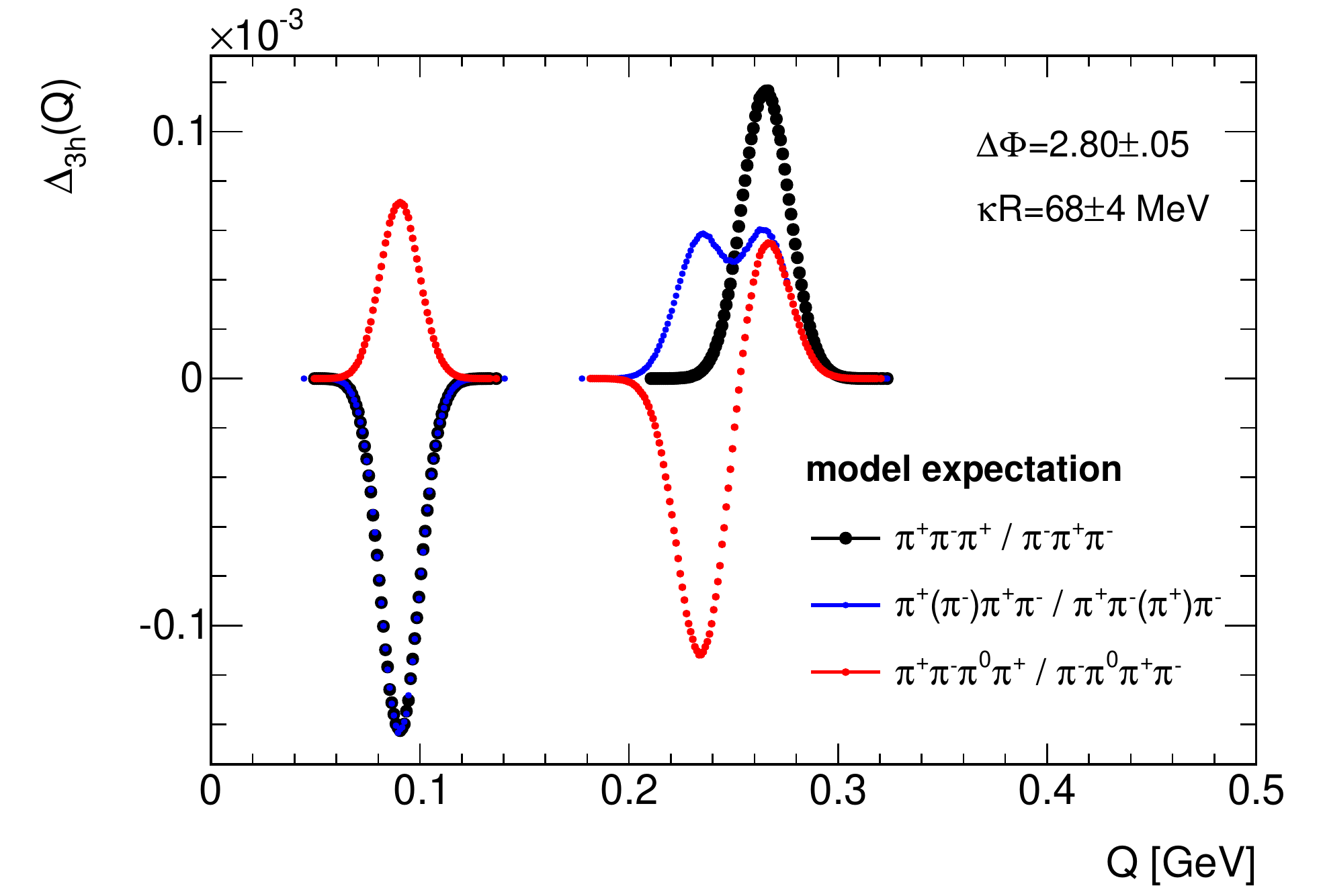}
\caption{ Evaluation of the model prediction for the shape of the correlation function $\Delta_{\mathrm 3h}$ 
  produced by the ground-state triplet (black points), ground-state quadruplet with missing
  middle member (blue points)  and ground-state quadruplet containing a
  neutral pion (red points), for equal number of triplets of each type.  Model predictions are calculated 
  using a Gaussian smearing of helical string parameters constrained by the fit of mass spectrum of 
  pseudoscalar mesons. The resolution uncertainty of the order of 10~MeV is included in the
  variation of string parameters. 
\label{fig:model_dq3h}}
\end{center}
\end{figure}

  The helix-string model predictions for triplets of charged pions composed of particles with rank difference $r$<4 are shown in Figure~\ref{fig:dalitz_exp}, using
 a numerical smearing of the measured parameters $\kappa R$ and $\Delta\Phi$ (by a Gaussian distribution with a width of  6\% and 2\%, respectively). Figure~\ref{fig:model_dq3h} shows the corresponding shapes of the correlation function $\Delta_{\mathrm 3h}(Q)$. The presence of a pair with rank difference 3 modifies
the shape of the positive part of the $\Delta_{\mathrm 3h}$ distribution. The presence of a neutral pion within the chain tends to diminish the correlation signal
by compensating, to some extent, the asymmetry between the like-sign and the opposite-sign pair $Q$ distributions.  

\printbibliography

\newpage \input{atlas_authlist}

\end{document}

%% file: atlas_authlist.tex
\begin{flushleft}
{\Large The ATLAS Collaboration}

\bigskip

M.~Aaboud$^\textrm{\scriptsize 137d}$,
G.~Aad$^\textrm{\scriptsize 88}$,
B.~Abbott$^\textrm{\scriptsize 115}$,
O.~Abdinov$^\textrm{\scriptsize 12}$$^{,*}$,
B.~Abeloos$^\textrm{\scriptsize 119}$,
S.H.~Abidi$^\textrm{\scriptsize 161}$,
O.S.~AbouZeid$^\textrm{\scriptsize 139}$,
N.L.~Abraham$^\textrm{\scriptsize 151}$,
H.~Abramowicz$^\textrm{\scriptsize 155}$,
H.~Abreu$^\textrm{\scriptsize 154}$
% ...
\bigskip
\\
% ...
$^{12}$ Institute of Physics, Azerbaijan Academy of Sciences, Baku, Azerbaijan\\
% ...
$^{88}$ CPPM, Aix-Marseille Universit{\'e} and CNRS/IN2P3, Marseille, France\\
% ...
$^{115}$ Homer L. Dodge Department of Physics and Astronomy, University of Oklahoma, Norman OK, United States of America\\
% ...
$^{119}$ LAL, Univ. Paris-Sud, CNRS/IN2P3, Universit{\'e} Paris-Saclay, Orsay, France\\
% ...
$^{137}$ $^{(a)}$ Facult{\'e} des Sciences Ain Chock, R{\'e}seau Universitaire de Physique des Hautes Energies - Universit{\'e} Hassan II, Casablanca; $^{(b)}$ Centre National de l'Energie des Sciences Techniques Nucleaires, Rabat; $^{(c)}$ Facult{\'e} des Sciences Semlalia, Universit{\'e} Cadi Ayyad, LPHEA-Marrakech; $^{(d)}$ Facult{\'e} des Sciences, Universit{\'e} Mohamed Premier and LPTPM, Oujda; $^{(e)}$ Facult{\'e} des sciences, Universit{\'e} Mohammed V, Rabat, Morocco\\
% ...
$^{139}$ Santa Cruz Institute for Particle Physics, University of California Santa Cruz, Santa Cruz CA, United States of America\\
% ...
$^{151}$ Department of Physics and Astronomy, University of Sussex, Brighton, United Kingdom\\
% ...
$^{154}$ Department of Physics, Technion: Israel Institute of Technology, Haifa, Israel\\
$^{155}$ Raymond and Beverly Sackler School of Physics and Astronomy, Tel Aviv University, Tel Aviv, Israel\\
% ...
$^{161}$ Department of Physics, University of Toronto, Toronto ON, Canada\\
% ...
$^{a}$ Also at Department of Physics, King's College London, London, United Kingdom\\
$^{b}$ Also at Institute of Physics, Azerbaijan Academy of Sciences, Baku, Azerbaijan\\
$^{c}$ Also at Novosibirsk State University, Novosibirsk, Russia\\
$^{d}$ Also at TRIUMF, Vancouver BC, Canada\\
$^{e}$ Also at Department of Physics {\&} Astronomy, University of Louisville, Louisville, KY, United States of America\\
% ...
$^{*}$ Deceased
\end{flushleft}
